\documentclass{aa}  

\usepackage{graphicx}
\usepackage[varg]{txfonts}
\usepackage{xcolor}
                                
\begin{document}

\title{Hydrodynamical simulations of the vertical shear instability with dynamic dust and cooling rates in protoplanetary disks}

\author{
    Y. Fukuhara\inst{1,2} \and 
    M. Flock\inst{3} \and
    S. Okuzumi\inst{2} \and
    R. T. Tominaga\inst{2}
    }

\institute{
    Institute of Astronomy and Astrophysics, Academia Sinica, 11F of Astronomy-Mathematics Building, No.1, Sec. 4, Roosevelt Rd, Taipei 106319, Taiwan, R.O.C.\\
    \email{yfukuhara@asiaa.sinica.edu.tw}
    \and Department of Earth and Planetary Sciences, Institute of Science Tokyo, Meguro, Tokyo 152-8551, Japan
    \and Max Planck Institute for Astronomy, K\"{o}nigstuhl 17, 69117 Heidelberg, Germany\\
    }



\abstract
{ 
Turbulence in protoplanetary disks affects dust evolution and planetesimal formation.
The vertical shear instability (VSI) is one of the candidate turbulence-driving mechanisms in the outer disk region.
Since the VSI requires rapid gas cooling, dust particles in disks can influence and potentially control VSI-driven turbulence.
However, VSI-driven turbulence has strong vertical motion, causing vertical diffusion of particles.
As a result of this interaction, it remains unclear how turbulent structures and dust distributions form and persist.
}
{ 
We aim to clarify whether the VSI can drive turbulence and achieve a quasi-steady dust distribution under cooling rate evolution associated with turbulently diffusing dust.
We also elucidate the dependence of the dust size and dust-to-gas mass ratio on the realization and persistence of the equilibrium state.
}
{ 
We perform global two-dimensional hydrodynamical simulations of an axisymmetric protoplanetary disk to investigate how the VSI drives turbulence and maintains a balance between dust settling and diffusion. 
These simulations account for the dynamic interplay between dust distribution, cooling rates, and turbulence.
}
{ 
We find that VSI mixing, dust settling, and local dust cooling reach an equilibrium, forming a thick dust layer with a dimensionless vertical mixing coefficient of approximately $\alpha_{\rm diff} \approx 10^{-3}$.
The ability of the VSI to sustain this state also depends on the dust size and dust-to-gas mass ratio. 
Larger grains or lower mass ratios weaken turbulence, leading to dust settling.
The condition of equilibrium state existence is consistent with the prediction of the semi-analytic model presented by \citet{FukuharaOkuzumi:2024aa}.
}
{ 
Our results indicate that efficient turbulent dust mixing and efficient cooling can occur simultaneously.
They also imply that turbulence in VSI-dominated disks has different levels of intensity depending on the grain size.
This suggests that the efficiency of dust growth can depend on the VSI in protoplanetary disks.
}


\titlerunning{Hydrodynamical simulations of the VSI with dynamic dust and cooling rates in protoplanetary disks}
\authorrunning{Y. Fukuhara et al.}

\maketitle

\nolinenumbers 

\section{Introduction}\label{sec:intro}

The initial stage of planet formation involves the growth of micrometer-sized dust particles into kilometer-sized planetesimals within a protoplanetary disk around a young star (for a review, \citealt{DrazkowskaBitsch+:2023aa}).
This dust evolution and planetesimal formation depend on gas turbulence in disks.
Gas turbulence can induce relative velocities between dust particles (e.g., \citealt{OrmelCuzzi2007}), leading to their fragmentation (e.g., \citealt{Brauer:2008aa,Okuzumi+2012}).
Turbulence can also transport dust particles and determine their spatial distribution (e.g., \citealt{Dubrulle+1995,UmurhanEstrada+:2020yi,RaettigLyra+:2021sb}).
Understanding the mechanisms that drive turbulence is essential for interpreting recent high-resolution radio observations, which suggest that the intensity of turbulence varies with radial and vertical location, as well as between different disks (for reviews, \citealt{MiotelloKamp+:2023aa,Rosotti:2023aa}).

The vertical shear instability (VSI) is one of the most robust candidates for driving turbulence in the outer parts of protoplanetary disks (for reviews, \citealt{LyraUmurhan2019,LesurFlock+:2023aa}).
The VSI requires a vertical gradient in rotational velocity (e.g., \citealt{NelsonGresselUmurhan2013}), which is called vertical shear, and rapid gas cooling (e.g., \citealt{LinYoudin2015,MangerPfeil+:2021cm}).
The VSI generates turbulence characterized by predominant vertical gas motions (e.g., \citealt{NelsonGresselUmurhan2013,StollKley2014,ShariffUmurhan:2024aa}), which can inhibit the settling of dust toward the midplane (e.g., \citealt{StollKley:2016vp,FlockNelson+2017,Flock:2020aa,DullemondZiampras+:2022aa,PfeilBirnstiel+:2024aa}).
Because gas molecules are less efficient radiators than dust particles (e.g., \citealt{Malygin+2017,BarrancoPei+:2018kc}), the size and spatial distribution of dust determine the spatial profile of the cooling rate \citep{PfeilKlahr2019,FukuharaOkuzumi+:2021ca,FukuharaOkuzumi+:2023aa}, thereby regulating VSI-driven turbulence \citep{PfeilKlahr:2021nr,PfeilBirnstiel+:2023aa,FukuharaOkuzumi+:2023aa}.
In outer disk regions that are optically thin to dust thermal emission, the VSI can operate due to efficient cooling by dust particles \citep{Malygin+2017,PfeilKlahr2019,FukuharaOkuzumi:2024aa}.
Even in optically thick regions, dust growth and settling can change the optical depth and trigger VSI-driven turbulence.

The previous study predicts that an equilibrium state between VSI-driven turbulence and the vertical distribution of dust grains exists for small grains but disappears for large ones \citep{FukuharaOkuzumi:2024aa}.
These results were obtained using a semi-analytic model where the dust settling balances with turbulent diffusion in a self-consistent manner (see their Fig. 1) and a time-dependent toy model where the vertical distribution of dust evolves through settling and diffusion [see their Eq. (26)].
In both models, it was assumed that the dust vertical distribution established by VSI-driven turbulence follows a Gaussian profile characterized by a single vertical diffusion coefficient.
This assumption corresponds to a uniform mixing of dust particles across the entire vertical region, including regions beyond those where the VSI-driven turbulence is active.
However, the gas motion driven by the VSI is confined to the VSI-active region, and its intensity varies with height \citep{FukuharaOkuzumi+:2023aa}.
Because dust particles are transported by time-variable turbulent gas motions, their spatial distribution evolves dynamically (e.g., \citealt{StollKley:2016vp,Flock:2020aa,PfeilBirnstiel+:2023aa}) and may deviate significantly from a Gaussian profile.
Moreover, \citet{FukuharaOkuzumi:2024aa} adopted an empirical formula for the intensity of VSI-driven turbulence, derived from the hydrodynamical simulations that assumed a constant spatial distribution of the cooling timescale (see \citealt{FukuharaOkuzumi+:2023aa}). 
In reality, the cooling timescale evolves on timescales comparable to that of turbulence because it depends on the dust distribution, which is itself modified by turbulent gas motion (e.g., \citealt{FukuharaOkuzumi+:2021ca,PfeilBirnstiel+:2023aa}).
Therefore, it remains unclear what kinds of dust distributions can emerge under VSI-driven turbulence, what turbulent structures can develop, and how such states can be maintained over time.

Recently, \citet{PfeilBirnstiel+:2024aa} performed hydrodynamic simulations with thermal accommodation times that are consistent with the simulated dust density and grain size.
Those simulations included dust growth evolution based on an analytical model, similar to that of \citet{Birnstiel:2012aa}.
They found that the VSI-driven turbulence can be maintained even as dust grows, and that achieving such a state requires an initially small grain and a large initial dust scale height.
However, dust growth itself should depend on turbulence, and the dust size distribution should also vary locally because the degree of gas--dust coupling is size-dependent.
In addition, how the dust-to-gas mass ratio, which influences the cooling rate, depends on the turbulence-driving conditions remains unclear.

In this study, we investigate how VSI-driven turbulence drives and maintains dust diffusion under dynamically evolving dust and cooling rates.
We perform global 2.5-dimensional (two dimensions for space and three dimensions for the velocity vector components) gas--dust hydrodynamical simulations of an axisymmetric protoplanetary disk with a cooling rate that evolves dynamically in response to dust density.
We show that the existence of quasi-steady equilibrium states, in which dust settling balances with VSI-driven turbulent diffusion, depends on the dust grain size and dust-to-gas mass ratio.

This paper is organized as follows. 
In Sect. \ref{sec:method}, we describe our numerical hydrodynamical simulation setup and simulation analysis method.
We present the main results in Sect. \ref{sec:results}, and discuss the applications in Sect. \ref{sec:discussion}.
Section \ref{sec:conclusions} presents the conclusions of this study.

\section{Method}\label{sec:method}
We perform global 2.5-dimensional two-fluid (gas and pressureless dust) hydrodynamic simulations of an axisymmetric protoplanetary disk in spherical polar coordinates $(r,~\theta,~\phi)$.
We also use the cylindrical radius $R=r\sin{\theta}$ and vertical coordinate $z=r\cos{\theta}$.
We consider only the drag force exerted by gas on dust and neglect the feedback from dust to gas. 
This approximation is valid as long as the local dust-to-gas mass ratio remains smaller than unity.
We discuss this in more detail in Sect. \ref{subsec:Limitation}. 
The solved continuity, motion, and energy equations of gas hydrodynamics are given by
\begin{equation}\label{eq:continuity_gas}
    \frac{\partial \rho_{\rm gas}}{\partial t} + \nabla\cdot (\rho_{\rm gas} \boldsymbol{v}_{\rm gas}) = 0,
\end{equation}
\begin{equation}\label{eq:motion_gas}
    \frac{\partial \rho_{\rm gas}\boldsymbol{v}_{\rm gas}}{\partial t} + \nabla\cdot \left(\rho_{\rm gas} \boldsymbol{v}_{\rm gas}\boldsymbol{v}_{\rm gas}^T\right) = -\nabla P - \rho_{\rm gas}\nabla \Phi,
\end{equation} 
\begin{equation}\label{eq:energy_gas}
    \frac{\partial E_{\rm t}}{\partial t} + \nabla\cdot[(E_{\rm t} +P)\boldsymbol{v}_{\rm gas}] = -\rho_{\rm gas}\boldsymbol{v}_{\rm gas}\cdot\nabla \Phi,
\end{equation}
where $\rho_{\rm gas}$ is the gas density, $\boldsymbol{v}_{\rm gas}$ is the gas velocity, $P$ is the gas pressure, $\Phi$ is the gravitational potential of the central star, and $E_{\rm t}$ is the total gas energy per unit volume.
The gravitational potential is given by $\Phi = -GM/R$, where $G$ is the gravitational constant and $M$ is the central star's mass.
In this study, $M$ is taken to be $1M_\odot$.
The total gas energy per unit volume is given by
\begin{equation}
    E_{\rm t} = \frac{P}{\gamma-1} + \frac{1}{2}\rho_{\rm gas}\boldsymbol{v}_{\rm gas}^2,
\end{equation}
where $\gamma$ is the heat capacity ratio and is taken to be 1.42.

The solved continuity and motion equations of pressureless dust are given by
\begin{equation}\label{eq:continuity_dust}
    \frac{\partial \rho_{\rm dust}}{\partial t} + \nabla\cdot (\rho_{\rm dust} \boldsymbol{v}_{\rm dust}) = 0,
\end{equation}
\begin{equation}\label{eq:motion_dust}
    \frac{\partial \rho_{\rm dust}\boldsymbol{v}_{\rm dust}}{\partial t} + \nabla\cdot \left(\rho_{\rm dust} \boldsymbol{v}_{\rm dust}\boldsymbol{v}_{\rm dust}^T\right)= - \rho_{\rm dust}\nabla \Phi +\rho_{\rm dust}\frac{\boldsymbol{v}_{\rm gas}-\boldsymbol{v}_{\rm dust}}{t_{\rm stop}},
\end{equation}
where $\rho_{\rm dust}$ is the dust density, $\boldsymbol{v}_{\rm dust}$ is the dust velocity, and $t_{\rm stop}$ is the stopping time. 

We assume the single-size dust where all particles are equal size $a$.
For the stopping time $t_{\rm stop}$, we assume that the size of the dust grains is smaller than the mean free path of the gas molecules and adopt
\begin{equation}\label{eq:stopping_time}
    t_{\rm stop} = \frac{\rho_{\rm int}a}{\rho_{\rm gas}v_{\rm th}},
\end{equation}
where $\rho_{\rm int}$ is the grain's internal density, and $v_{\rm th}=\sqrt{8/\pi}c_{\rm s}$ is the gas thermal velocity with $c_{\rm s}$ being the sound speed.
Equation \eqref{eq:stopping_time} satisfies the condition that the stopping time complies with Epstein's law.
In this study, we fix the dust internal density $\rho_{\rm int}$ to $1.46~{\rm g~cm^{-3}}$.
The gas and dust velocities have three components for the radial, meridional, and azimuthal velocities in spherical polar coordinates, $\boldsymbol{v}_{\rm gas} = (v_{r, \rm gas},~v_{\theta,\rm gas},~v_{\phi,\rm gas})$ and $\boldsymbol{v}_{\rm dust} = (v_{r, \rm dust},~v_{\theta,\rm dust},~v_{\phi,\rm dust})$.

For the cooling (thermal relaxation) model, we adopt the $\beta$ cooling model as \citep{NelsonGresselUmurhan2013,MangerKlahr:2018dw,PfeilKlahr:2021nr}
\begin{equation}\label{eq:cooling_equation}
    \frac{\partial P}{\partial t} = -\frac{\rho_{\rm gas}}{\gamma}\frac{c_{\rm s}^2-c_{\rm s, ini}^2}{t_{\rm cool}},
\end{equation}
where $c_{\rm s, ini}$ and $t_{\rm cool}$ are the initial isothermal sound speed and local cooling timescale, respectively.
Here, we use the ideal equation of state as $P=\rho_{\rm gas}c_{\rm s}^2/\gamma$.
Assuming that gas density and cooling timescale are constant in both time and space during the time step, Eq. \eqref{eq:cooling_equation} has a simple analytic solution of the form
\begin{equation}\label{eq:cooling_equation_analytic}
    P^{(n+1)} = \frac{1}{\gamma}\rho_{\rm gas}^{(n)}c_{\rm s,ini}^2+\frac{1}{\gamma}\rho_{\rm gas}^{(n)}\left[\left\{c_{\rm s}^{(n)}\right\}^2-c_{\rm s,ini}^2\right]\exp{\left(-\frac{\Delta t}{t_{\rm cool}}\right)},
\end{equation}
where $P^{(n)}$, $\rho_{\rm gas}^{(n)}$, and $c_{\rm s}^{(n)}$ are the pressure, gas density, and sound speed at $n$ time steps, respectively, and $\Delta t$ is the time step.
We calculate Eq. \eqref{eq:cooling_equation_analytic} at every time step as cooling.
In this study, the cooling timescale depends on the dust density and dust grain size.
We describe more details of the $t_{\rm cool}$ model in Sect. \ref{subsec:cooling_timescale_model}.

To solve the hydrodynamical equations [Eqs. \eqref{eq:continuity_gas}--\eqref{eq:motion_dust}], we use the PLUTO code \citep{MignoneBodo+:2007aa} including the pressureless dust module \citep{ZiamprasSudarshan+:2025aa} with the combination of the reconstruction scheme of a forth-order piecewise parabolic method \citep{Mignone:2014aa} and third-order Runge--Kutta time integrator.
We employ the Harten--Lax--van Leer--contact (HLLC) approximate Riemann solver \citep{MignoneBodo:2005bv} for equations of gas [Eqs. \eqref{eq:continuity_gas}--\eqref{eq:energy_gas}] and the exact Riemann solver \citep{Leveque:2004aa} for equations of pressureless dust [Eqs. \eqref{eq:continuity_dust}--\eqref{eq:motion_dust}].
The Courant--Friedrichs-Lewy (CFL) number is set to 0.25.

    \subsection{Simulation setup}\label{subsec:simulation_setup}
    
    \begin{table}[t]
    \begin{center}
    \caption{Setup parameters for global hydrodynamics simulations}
    \begin{tabular}{lcc} 
    \hline \hline
    Parameter & Symbol & Value \\ \hline
    Reference radius & $R_0$ & $100~{\rm au}$ \\ 
    Reference gas column density & $\Sigma_0$ & $6~{\rm g~cm^{-2}}$ \\
    Reference gas scale height & $H_0$ & $10~{\rm au}$ \\
    Radial power–low index of gas density & $p$ & -1 \\
    Radial power–low index of temperature & $q$ & -1 \\
    \hline
    \end{tabular} \\
    \label{table:setup_parameter}
    \end{center}
    \end{table}

    For the initial condition, the spatial profile of the gas density is given by (e.g., \citealt{NelsonGresselUmurhan2013})
    \begin{equation}\label{eq:gas_density_profile}
        \rho_{\rm gas}(R,~z) = \rho_0\left(\frac{R}{R_0}\right)^p\exp{\left[\left(\frac{R}{\sqrt{R^2+z^2}}-1\right)\left(\frac{R}{H_{\rm gas}}\right)^2\right]},
    \end{equation}
    where $\rho_0$ is the reference gas density, $R_0$ is the reference radius, $p$ is the radial power-law index for the gas density, and $H_{\rm gas}$ is the gas scale height.
    The gas scale height is given by
    \begin{equation}
        H_{\rm gas} = \frac{c_{\rm s}}{\Omega_{\rm K}}=H_0\left(\frac{R}{R_0}\right)^{(q+3)/2},
    \end{equation}
    where $\Omega_{\rm K} = \sqrt{GM/R^3}$ is the Keplerian frequency, $q$ is the radial power-law index for the temperature, and $H_0$ is the gas scale height at the reference radius.
    To evaluate the reference gas density, we introduce the gas surface density at the reference radius defined by $\Sigma_0 = \int \rho_{\rm gas}(R_0,~z) dz$.
    Because this integral is dominated by the region $z\lesssim H_{\rm gas}$, Eq. \eqref{eq:gas_density_profile} can be approximated as $\rho_{\rm gas}(R_0,~z) \approx \rho_0\exp(-z^2/(2H_0^2))$, yielding 
    \begin{equation}\label{eq:reference_gas_density}
        \rho_0 = \frac{\Sigma_0}{\sqrt{2\pi}H_0}.
    \end{equation}
    Our parameter choices are summarized in Table \ref{table:setup_parameter}.

    For the velocity components, the initial values are set to $v_{r,\rm gas} = v_{r, \rm dust} = v_{\theta,\rm gas} = v_{\theta,\rm dust} = 0$ and $v_{\phi,\rm gas} = R\Omega(R,~z)$, where $\Omega(R,~z)$ is the gas angular velocity.
    Assuming the radial and vertical force balance and radial gas density and temperature profile following power law, $\Omega(R,~z)$ is given by \citep{TakeuchiLin2002,NelsonGresselUmurhan2013}
    \begin{equation}
        \Omega(R,~z) = \Omega_{\rm K}\left[1+q\left(1-\frac{R}{r}\right)+\left(p+q\right)\left(\frac{H_{\rm gas}}{R}\right)^2\right]^{1/2}.
    \end{equation}
    For the azimuth velocity component of the dust, we assume that the dust is perfectly relaxed in the gas and set $v_{\phi,\rm dust} = v_{\phi,\rm gas}$.
    We also add small cellwise random velocities with an amplitude of $10^{-5}c_{\rm s}$ to the initial gas radial and meridional velocity field.

    For the dust density profile, we assume that the dust is completely mixed in the early stage of disk evolution.
    This assumption is consistent with the temporal trend in dust scale heights estimated by recent disk observations (e.g., \citealt{OhashiTobin+:2023aa,VillenavePodio+:2023aa,LinLi+:2023aa,EncaladaLooney+:2024aa}).
    The initial dust density distribution can be expressed as 
    \begin{equation}\label{eq:Z_d2g_initial}
        \rho_{\rm dust}(R,~z) = Z_{\rm D2G}\rho_{\rm gas}(R,~z),
    \end{equation}
    where $Z_{\rm D2G}$ is the initial dust-to-gas mass ratio.
    This equation yields that the dust scale height $H_{\rm dust}$ is equal to the gas scale height in the initial condition.

    For the gas velocity components and all dust components, we apply the outflow boundary condition, preventing inflow at the radial and meridional boundaries.
    At the inner and outer radial boundaries, we fix both gas density and pressure to the initial values.
    At the upper and lower meridional boundaries, we use logarithmic-continuous vertical gradient conditions for the gas density and vertically isothermal conditions for the gas pressure.

    The radial computational domain has a range with logarithmically spaced grid cells and a range with stretched grid cells.
    The meridional grid cells have linearly uniform spacing.
    The radial and meridional domains cover $20~{\rm au}\leq r \leq 150~{\rm au}$, with $20~{\rm au}\leq r \leq 100~{\rm au}$ allocated to the logarithmically spaced zone and $100~{\rm au}\leq r \leq 150~{\rm au}$ allocated to the stretch zone, and $\pi/2-0.35 \leq \theta \leq \pi/2+0.35$, respectively.
    Except for the stretched zone, we adopt a resolution of around $70$ cells per gas scale height in both radial and meridional direction, yielding cells with an aspect ratio of approximately $1:1$.
    This resolution is sufficient to properly analyze the global structure and intensity of VSI-driven turbulence \citep{Flores-Rivera:2020ab}\footnote{Because large-scale gas motion determines dust settling and vertical diffusion, the smaller vortex structures shown in the higher resolution simulations \citep{Flores-Rivera:2020ab,CuiLatter:2022aa,Melon-FuksmanFlock+:2024ab} would not have a significant effect on the dust vertical distribution.}.
    Therefore, our simulations use $1120\times 512$ grid cells, including $96$ cells at the stretched zone in the radial direction.

    In addition, we set the buffer zone at $20~{\rm au}\leq r \leq 23~{\rm au}$ and $120~{\rm au}\leq r \leq 150~{\rm au}$ to suppress the effect of wave reflection at the boundaries.
    Within the buffer zones, we damp the gas density, pressure, and gas and dust velocities to the initial values and cooling timescale to $100\Omega_{\rm K}^{-1}$ that corresponds to the adiabatic state.
    For the damping zone, we use the form \citep{de-Val-BorroEdgar+:2006aa,MangerKlahr:2018dw,HuangBai:2022aa}
    \begin{equation}
        \frac{\partial\boldsymbol{x}}{\partial t} = -\frac{\boldsymbol{x}-\boldsymbol{x}_0}{\tau_{\rm damp}}\cdot\left(\frac{r-r_{\rm damp}}{w_{\rm damp}}\right)^2,
    \end{equation}
    where $\boldsymbol{x}$ represent the relaxed physical quantities, $\boldsymbol{x}_0$ is the value at the computational boundaries for each physical quantity, $\tau_{\rm damp}$ is the damping timescale, $r_{\rm damp}$ is the radial edge of the buffer zone, and $w_{\rm damp}$ is the radial width of the buffer zone.
    The parabolic function $[(r-r_{\rm damp})/w_{\rm damp}]^2$ smoothly changes from zero at the buffer zone edge to unity at the boundary.
    We adopt $\boldsymbol{x} = (\rho_{\rm gas},~P_{\rm gas},~\boldsymbol{v}_{\rm gas}, ~\boldsymbol{v}_{\rm dust},~t_{\rm cool} )^T$ for the inner buffer zone and $\boldsymbol{x} = (\rho_{\rm gas},~P_{\rm gas},~\boldsymbol{v}_{\rm gas})^T$ for the outer buffer zone.
    For $\boldsymbol{x_0}$, the cooling timescale $t_{\rm cool}$ is set to $100\Omega_{\rm K}^{-1}$, otherwise to the initial values.
    We set the damping timescale $\tau_{\rm damp}$ to the time step $\Delta t$ for all physical quantities in the inner and outer buffer zones.
    For the inner and outer buffer zones, we set ($r_{\rm damp},~w_{\rm damp}$) to ($23~{\rm au},~3{\rm ~au}$) and ($120~{\rm au},~30~{\rm au}$), respectively.

    Furthermore, we set the drag force from gas to dust $\boldsymbol{f}_{\rm drag}= -\rho_{\rm dust}(\boldsymbol{v}_{\rm gas}-\boldsymbol{v}_{\rm dust})/t_{\rm stop}$ to damp in the inner buffer zone.
    We apply the parabolic function to satisfy the condition of $\boldsymbol{f}_{\rm drag} = 0$ at the computational inner radial boundary and $\boldsymbol{f}_{\rm drag} = \boldsymbol{f}_{\rm drag}|_{r=r_{\rm damp}}$ at the edge of the inner buffer zone and smoothly change, yielding
    \begin{equation}
        \boldsymbol{f}_{\rm drag} = -\boldsymbol{f}_{\rm drag}|_{r=r_{\rm damp}} \left(\frac{r-r_{\rm damp}}{w_{\rm damp}}\right)^2 + \boldsymbol{f}_{\rm drag}|_{r=r_{\rm damp}}.
    \end{equation}

    For time unit, we use the orbital period $P_{\rm in}$ at $R=20~{\rm au}$, defined by $P_{\rm in} = 2\pi/\Omega_{\rm K}(R=20~{\rm au})$. 

    \subsection{Cooling timescale}\label{subsec:cooling_timescale_model}
    Because the VSI requires rapid gas cooling, the spatial cooling rate profile, as well as its values, determines where the linear growth of the VSI occurs.
    Following \citet{Malygin+2017} and \citet{FukuharaOkuzumi+:2021ca}, we use the linear instability criterion, which can be expressed as
    \begin{equation}\label{eq:VSI_criterion}
        t_{\rm cool} \lesssim t_{\rm crit},
    \end{equation}
    where 
    \begin{equation}\label{eq:t_crit}
        t_{\rm crit} = \frac{|q|}{\gamma-1}\frac{H_{\rm g}}{R}\Omega_{\rm K}^{-1}
    \end{equation}
    is the global critical cooling time \citep{LinYoudin2015}.
    In this study, we refer to regions that fulfill this criterion as the VSI-active regions\footnote{\citet{FukuharaOkuzumi+:2021ca} and \citet{FukuharaOkuzumi:2024aa} called these the VSI zones and VSI-unstable layers, respectively.}.
    The VSI-active regions exist on the upper ($z>0$) and lower ($z<0$) halves of the disk.
    For the upper halves, we determine the height of the VSI-active region's upper boundary $z_{\rm VSI}$. 
    Following \citet{FukuharaOkuzumi+:2023aa}, we define the thickness of the VSI-active regions as $\Delta L_{\rm VSI} = 2z_{\rm VSI}$.

    The cooling timescale in Eq. \eqref{eq:cooling_equation} depends on the dust spatial profile and grain size.
    In outer disk regions that are optically thin to their own thermal emission, the timescales of collisional heat transfer dominate the cooling timescale \citep{Malygin+2017,BarrancoPei+:2018kc,PfeilKlahr2019,FukuharaOkuzumi+:2021ca}.
    In this case, the cooling timescale can be expressed as\footnote{We note that this timescale differs by a factor of $\gamma/(\gamma-1)$ from the actual timescale of heat transport between gas molecules and dust particles \citep{BurkeHollenbach:1983aa,BarrancoPei+:2018kc,PfeilBirnstiel+:2023aa}.}
    \begin{equation}\label{eq:t_cool}
        t_{\rm cool} = \frac{\ell_{\rm gd}}{v_{\rm re}},
    \end{equation}
    where $\ell_{\rm gd}$ is the mean travel length of gas molecules before colliding with dust particles and $v_{\rm re}$ is the mean relative velocity between the gas molecules and dust particles.
    The relative velocity $v_{\rm re}$ can be approximated as the mean thermal speed of the gas molecule $v_{\rm th}$; therefore, $v_{\rm re} = v_{\rm th}$.
    For $\ell_{\rm gd}$, we assume a single-size dust, yielding
    \begin{equation}\label{eq:lgd}
        \ell_{\rm gd} = \left(\pi a^2 \frac{\rho_{\rm dust}}{m_{\rm dust}}\right)^{-1},
    \end{equation}
    where $m_{\rm dust}=(4\pi/3) \rho_{\rm int}a^3$ is the mass of the particle. 
    Therefore, the cooling timescale is a function of dust density and dust grain size, with dependence on $t_{\rm cool} \propto \rho_{\rm dust}^{-1}a$. 
    In the simulations, we take $t_{\rm cool}\Omega_{\rm K}$ not below $0.01$ and above $100$ to avoid tiny and huge values of $t_{\rm cool}\Omega_{\rm K}$ at the midplane and high $|z|$, respectively.

    \begin{figure}[t]
        \begin{center}
        \includegraphics[width=\hsize,bb = 0 0 527 323]{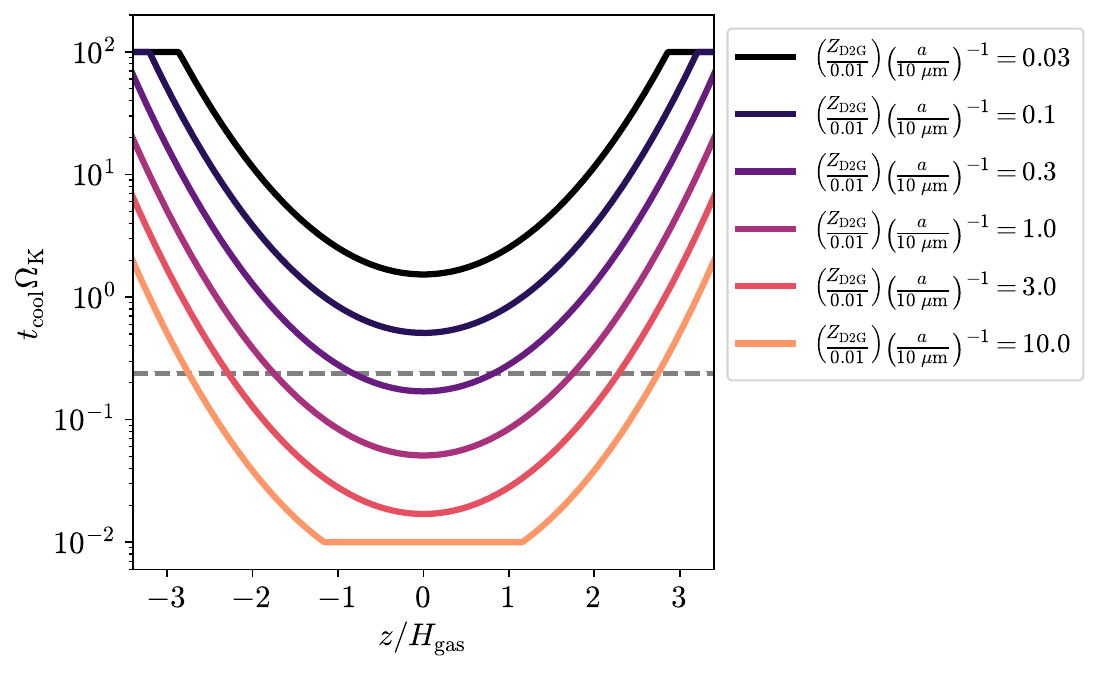}
        \end{center}
        \caption{Initial vertical profile of the cooling timescale $t_{\rm cool}\Omega_{\rm K}$ at $R=40~{\rm au}$ for different values of $(Z_{\rm D2G}/0.01)(a/10~{\rm \mu m})^{-1}$. The dashed line marks the critical cooling timescale in Eq. \eqref{eq:t_crit}.}
        \label{fig:vetical_profile_cooling_ini}
    \end{figure}  

    For the initial conditions of Eq. \eqref{eq:Z_d2g_initial} where dust is completely mixed with gas, the initial spatial profile of $t_{\rm cool}$ can be scaled by $Z_{\rm D2G}a^{-1}$.
    We plot in Fig. \ref{fig:vetical_profile_cooling_ini} the initial vertical profiles of $t_{\rm cool}$ at $R=40~{\rm au}$.
    This figure indicates that the runs with the combination of $Z_{\rm D2G}$ and $a$ satisfying $(Z_{\rm D2G}/0.01)(a/10~{\rm \mu m})^{-1} \geq 0.3$ have the VSI-active region around the midplane.

    \subsection{Two types of simulations and parameter choices}\label{subsec:parameter_choices}
    We perform two types of hydrodynamic simulations: (i) calculations with dust distribution and cooling timescale evolved from the beginning of the simulation, and (ii) calculations with cooling timescale fixed in time and space until turbulence saturates to a quasi-steady state, followed by dynamic dust and cooling timescale activation. 
    We call the former and latter ones "dynamic-dust" simulations and "fixed-to-dynamic-dust" simulations, respectively.
    
    The cooling timescale $t_{\rm cool}$ depends on the dust grain size $a$ and dust density $\rho_{\rm dust}$ [Eqs. \eqref{eq:t_cool} and \eqref{eq:lgd}].
    Therefore, the free parameters in our simulations are the dust grain size $a$ and initial local dust-to-gas mass ratio $Z_{\rm D2G}$.
    We take $a=10~{\rm \mu m}$, $30~{\rm \mu m}$, $100~{\rm \mu m}$, and $300~{\rm \mu m}$ and $Z_{\rm D2G} = 0.001$, $0.003$, $0.01$, $0.03$, and $0.1$.
    As mentioned in Sect. \ref{subsec:cooling_timescale_model}, runs with the same value of $Z_{\rm D2G}a^{-1}$ have the same initial distribution of cooling timescales.

    \subsection{Turbulence diagnostics}\label{subsec:turbulence_diagnostics}

    We quantify the dust vertical diffusion by VSI-driven turbulence using the dust layer thickness.
    To evaluate the dimensionless dust vertical diffusion coefficient $\alpha_{\rm diff}$, we estimate the dust layer thickness $2H_{\rm dg}$ from the vertical profile of $\rho_{\rm dust}/\rho_{\rm gas}$.
    We define $2H_{\rm dg}$ as the thickness of the region where $\rho_{\rm dust}/\rho_{\rm gas}$ is greater than half of the initial local dust-to-gas mass ratio $Z_{\rm D2G}$.
    In this study, the initial value of $2H_{\rm dg}$ equals the width of the computational meridional domain because we assume a spatially uniform $\rho_{\rm dust}/\rho_{\rm gas}$ in the initial condition [Eq. \eqref{eq:Z_d2g_initial}].
    We also calculate the dust scale height $H_{\rm dust}$, assuming that the vertical profile of $\rho_{\rm dust}$ within the dust layer follows a Gaussian profile of the form $\rho_{\rm dust,0}\exp{(-z^2/2H_{\rm dust}^2)}$, where $\rho_{\rm dust,0}$ is the dust density at the midplane.

    Assuming that the scale height of the "dust concentration" in the gas is equivalent to the dust layer's thickness, we can estimate the dust vertical turbulent diffusion coefficient $\alpha_{\rm diff}$ from $H_{\rm dg}$.
    Following \citet{Dubrulle+1995}, the scale height of the dust concentration can be expressed by
    \begin{equation}\label{eq:H_dg}
        H_{\rm dg} = \sqrt{\frac{\alpha_{\rm diff}}{{\rm St}}}H_{\rm gas},
    \end{equation}
    where ${\rm St} = t_{\rm stop}\Omega_{\rm K}$ is the dimensionless stopping timescale (Stokes number).
    When we estimate $\alpha_{\rm diff}$ using Eq. \eqref{eq:H_dg}, we use the value of ${\rm St}$ at the midplane.
    In this study, the dust sizes of $a=10~{\rm \mu m}$, $30~{\rm \mu m}$, $100~{\rm \mu m}$, and $300~{\rm \mu m}$ correspond to ${\rm St}\approx 7\times 10^{-4},~2\times 10^{-3},~7\times 10^{-3}$, and $2\times 10^{-2}$, respectively, at the midplane.
    Using Eq. \eqref{eq:H_dg}, we directly estimate the dust vertical diffusion coefficient $\alpha_{\rm diff}$ from the spatial profile of the dust density.
    Assuming that the dust vertical profile follows the turbulent vertical mixing-settling balance. the correlation between $H_{\rm dg}$ and $H_{\rm dust}$ can be expressed as \citep{Dubrulle+1995},
    \begin{equation}\label{eq:relation_Hdg_Hd}
        \frac{1}{H_{\rm dg}^2} = \frac{1}{H_{\rm dust}^2} + \frac{1}{H_{\rm gas}^2}.
    \end{equation}

    We quantify the strength of VSI-driven turbulence using the time average of the squared vertical velocity $\langle v_{z,\rm gas}^2\rangle$, where $v_{z,\rm gas} = v_{r,\rm gas}\cos{\theta} - v_{\theta,\rm gas}\sin{\theta}$ is the gas vertical velocity.
    The bracket $\langle \cdot \rangle$ denotes the time average.
    In our simulations, the time averaging is performed after the system relaxes into a quasi-steady state.
    To see the vertical global effect of turbulence, we also calculate the vertical mean of $v_{z,\rm gas}^2$ within the dust layer, which is denoted by overbars.

    The time that it takes for the system to reach a quasi-steady state differs from one run to another and may depend on the dust grain size and initial dust-to-gas mass ratio. 
    Therefore, in the cases of dynamic-dust simulations, we stop a run at $1000P_{\rm in}$ if the quasi-steady state has already been reached by $500P_{\rm in}$; otherwise, we continue the run until $2000P_{\rm in}$.
    In the fixed-to-dynamic-dust simulations, we first stop a run at $500P_{\rm in}$, using a fixed spatial profile for the cooling timescale.
    Next, we continue the simulations with a dynamic cooling timescale and stop the run at $500P_{\rm in}$ after the activation time of the dynamic cooling timescale, except in runs with $(Z_{\rm D2G},~a) = (0.01,~100~{\rm \mu m})$ and $(0.1,~10~{\rm \mu m})$, where we stop at $1500P_{\rm in}$.
    
\section{Results}\label{sec:results}
This section presents our simulation results to study what kind of dust spatial profile and VSI-driven turbulence structure can be achieved.
We show in Sect. \ref{subsec:equilibrium_state_laminar} the properties of the equilibrium states and the dependence of dust grain size and dust-to-gas mass ratio on it in dynamic-dust simulations.
In Sect. \ref{subsec:equilibrium_state_turbulent}, we present results in fixed-to-dynamic-dust simulations and show the condition of dust grain size and dust-to-gas mass ratio for dust diffusion by VSI-driven turbulence.

    \subsection{Equilibrium state with dynamic-dust simulations}\label{subsec:equilibrium_state_laminar}
    \begin{figure*}[t]
        \begin{center}
        \includegraphics[width=\hsize,bb = 0 0 834 301]{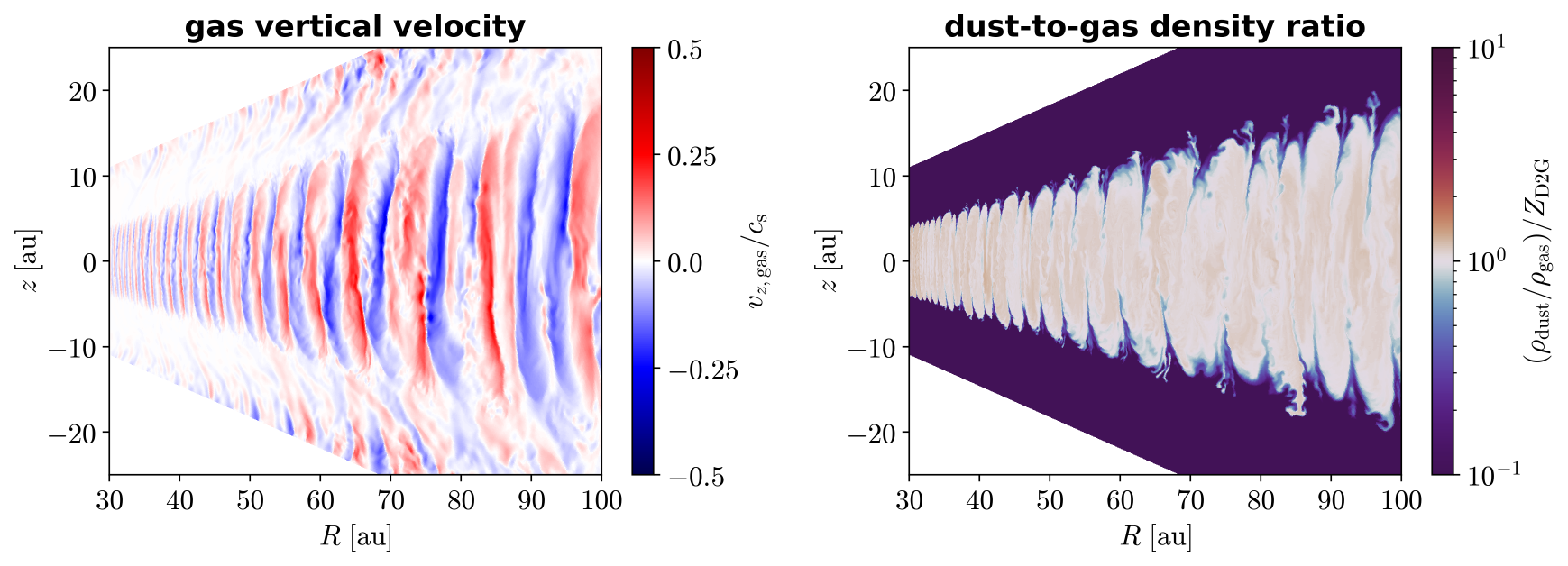}
        \end{center}
        \caption{Gas vertical velocity $v_{z,\rm gas}/c_{\rm s}$ (left) and ratio of dust and gas density $(\rho_{\rm dust}/\rho_{\rm gas})/Z_{\rm D2G}$ (right) for a run with $(Z_{\rm D2G},~a) = (0.01,~10~{\rm \mu m})$ in the dynamic-dust simulations, as a function of $R$ and $z$ at the end of the simulation ($1000$ orbits at $20~{\rm au}$).}
        \label{fig:colormap_vz_rhodgINI_D2G1e-2_a1e-3_h}
    \end{figure*}  

    We find that equilibrium-saturated quasi-steady states of VSI-driven turbulence and dust vertical profile exist.
    Figure \ref{fig:colormap_vz_rhodgINI_D2G1e-2_a1e-3_h} illustrates the equilibrium state.
    Here, we present two-dimensional maps of the gas vertical velocity $v_{z,\rm gas}$ normalized by the sound speed $c_{\rm s}$ and the ratio between dust density $\rho_{\rm dust}$ and gas density $\rho_{\rm gas}$ normalized by the initial global dust-to-gas mass ratio $Z_{\rm D2G}$, at the end of the simulation for a run with $(Z_{\rm D2G},~a) = (0.01,~10~{\rm \mu m})$.
    The vertically uniform gas motion is confined in the region with $|z|\lesssim 1.5 H_{\rm gas}$, corresponding to the dust layer.
    At higher altitudes, one can not see the vertical velocity structure caused by the VSI.
    The dust concentrates slightly $(\rho_{\rm dust}/\rho_{\rm gas})/Z_{\rm D2G} \approx 1.1$ in the region that exhibits the turbulence structure, while the dust is depleted $(\rho_{\rm dust}/\rho_{\rm gas})/Z_{\rm D2G} \ll 0.1$ in the upper layer due to settling to the dust layer.
    The fact that $(\rho_{\rm dust}/\rho_{\rm gas})/Z_{\rm D2G}$ remains close to unity in the dust layer indicates a combination of significantly efficient vertical mixing of the dust and efficient cooling at the same time in this regime.
    Figure \ref{fig:Z_saturated_mid} in appendix \ref{appendix:dust_consentration} presents the dust-to-gas density ratio at the midplane for runs where the equilibrium state exists.
    The spatial shape of the dust layer is associated with the spatial structure of the gas vertical velocity, which is similar to the structures in \citet{DullemondZiampras+:2022aa}.

    Our simulations do not clearly reproduce small eddies caused by the Kelvin--Helmholtz instability (KHI) or a finer spatial profile of vertical velocity caused by the secondary parametric instability, which can be related to the nonlinear saturation process of the VSI. 
    This is because our simulation resolution is insufficient to identify the eddies by the KHI or to resolve the parametric instability, which may require $\sim 100$ cells per gas scale height \citep{Melon-FuksmanFlock+:2024ab} or $\sim 300$ cells per gas scale height \citep{CuiLatter:2022aa}, respectively.

    \begin{figure*}[t]
        \begin{center}
        \includegraphics[width=\hsize,bb = 0 0 826 272]{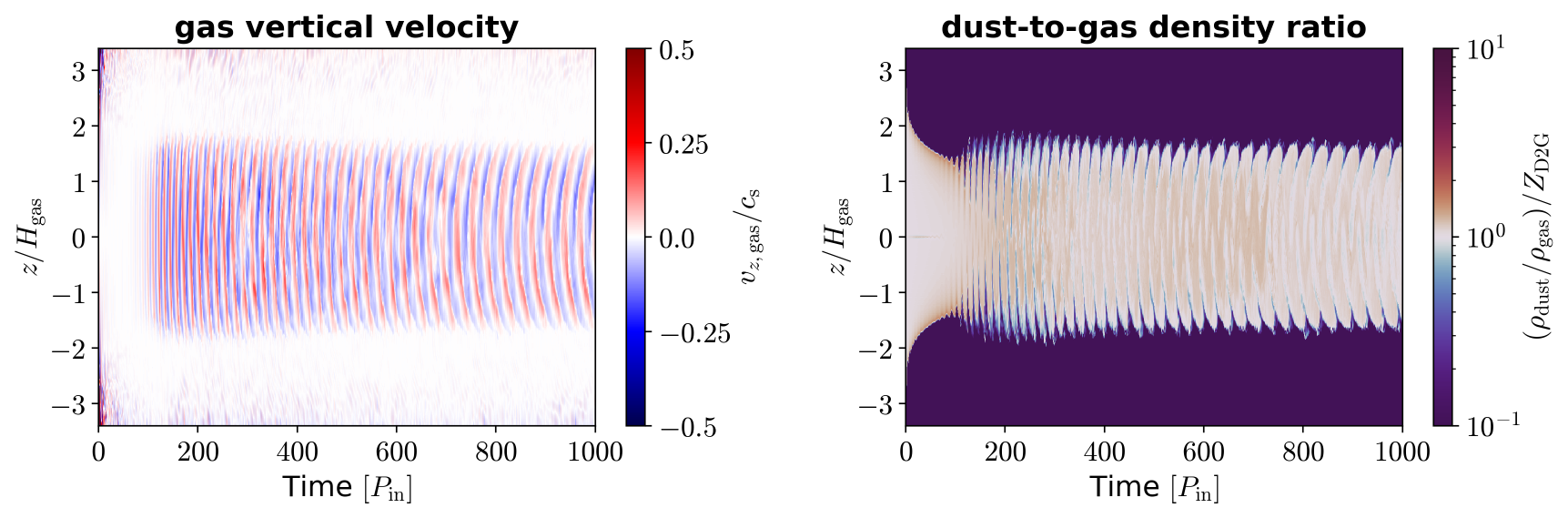}
        \end{center}
        \caption{Gas vertical velocity $v_{z,\rm gas}/c_{\rm s}$ (left) and ratio of dust and gas density $(\rho_{\rm dust}/\rho_{\rm gas})/Z_{\rm D2G}$ (right) for a run with $(Z_{\rm D2G},~a) = (0.01,~10~{\rm \mu m})$ in the dynamic-dust simulations, as a function of time and $z/H_{\rm gas}$ at $R=40~{\rm au}$.}
        \label{fig:colormap_time_vzINI_rhodg_D2G1e-2_a1e-3_h}
    \end{figure*}  

    To see how the final saturated states are reached, we plot in Fig. \ref{fig:colormap_time_vzINI_rhodg_D2G1e-2_a1e-3_h} the vertical profile of $v_{z,\rm gas}$ and $\rho_{\rm dust}/\rho_{\rm gas}$ at $R=40~{\rm au}$ for the case displayed in Fig. \ref{fig:colormap_vz_rhodgINI_D2G1e-2_a1e-3_h}.
    Before VSI-driven turbulence develops ($t\lesssim 200P_{\rm in}$), the dust settles toward the midplane, decreasing the dust layer thickness.
    As vertical gas motion begins to grow ($t\gtrsim 200P_{\rm in}$), it causes the dust to diffuse vertically to a height of $|z|\approx 1.5H_{\rm gas}$.
    Because the dust layer spread by vertical diffusion creates a thick VSI-active layer with a high cooling rate, vertical gas motion caused by the VSI sustains and continues to diffuse the dust.
    These structures are steady over $t=500$--$1000$ orbits, implying that our simulation captures the final saturated quasi-steady state.
    In this state, VSI-driven turbulence also dominates the dust radial motion (see appendix \ref{appendix:dust_radial_velocity}).

    \begin{figure}[t]
        \begin{center}
        \includegraphics[width=\hsize,bb = 0 0 434 659]{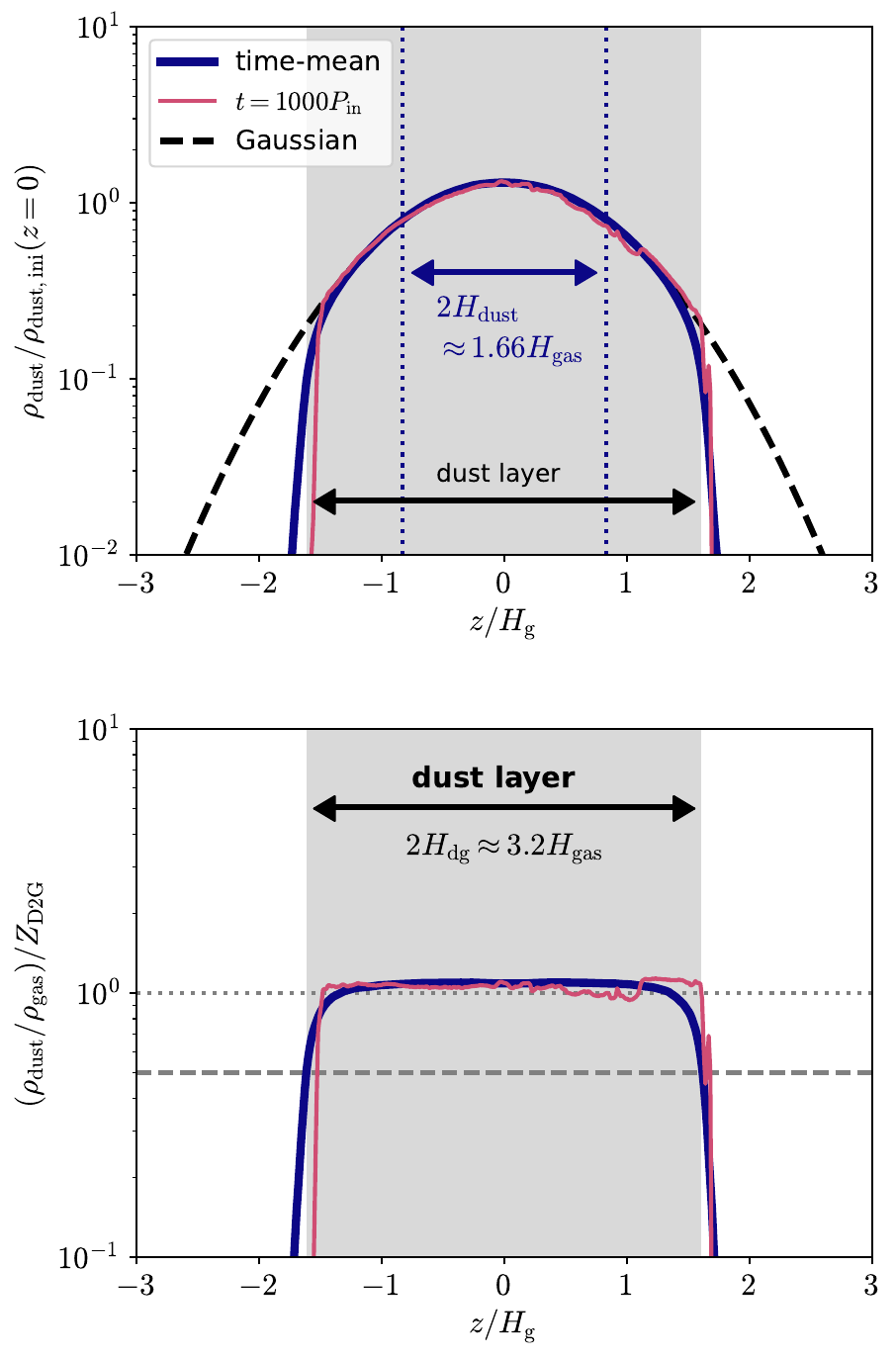}
        \end{center}
        \caption{Vertical profiles of the dust density $\rho_{\rm dust}$ normalized by the initial dust density at the midplane $\rho_{\rm dust,ini}(z=0)$ (upper panel), and ratio of dust and gas densities $\rho_{\rm dust}/\rho_{\rm gas}$ normalized by the initial dust-to-gas mass ratio $Z_{\rm D2G}$ (lower panel), at $R=40~{\rm au}$ for a run with $(Z_{\rm D2G},~a) = (0.01,~10~{\rm \mu m})$ in the dynamic-dust simulations. In the upper panel, the dashed curve shows a Gaussian fitting of $\rho_{\rm dust}$ within $|z|<H_{\rm dg}$, and the vertical dotted lines mark the height of the dust scale height $H_{\rm dust}$. The horizontal dashed and dotted lines in the lower panel mark $\rho_{\rm dust}/\rho_{\rm gas}=Z_{\rm D2G}/2$ and $Z_{\rm D2G}$, respectively. The gray regions in both panels map the dust layer defined by $H_{\rm dg}$.}
        \label{fig:vertical_profile_rhod}
    \end{figure} 

    The vertical distribution of the dust density deviates from the commonly assumed Gaussian distribution.
    The upper panel of Fig. \ref{fig:vertical_profile_rhod} plots the vertical profile of the dust density in the equilibrium state at $R=40~{\rm au}$.
    Within the region of $|z| \lesssim 1.6 H_{\rm gas}$, where VSI-driven turbulence is active, the dust density approximately follows a Gaussian profile.
    At higher altitudes, however, the dust density drops sharply due to the absence of turbulence, which otherwise counteracts settling.
    As a result, a dust layer forms within the turbulence-active region, where dust is more concentrated than in the initial state.
    The lower panel of Fig. \ref{fig:vertical_profile_rhod} shows $\rho_{\rm dust}/\rho_{\rm gas}$, normalized by $Z_{\rm D2G}$ at $R = 40~{\rm au}$.
    This profile indicates a relatively flat distribution of $\rho_{\rm dust}/\rho_{\rm gas}$, with values slightly exceeding the initial value $Z_{\rm D2G}$.
    Under this condition, the half-thickness of the dust layer and the dust scale height are approximately $H_{\rm dg} \approx 1.6 H_{\rm gas}$ and $H_{\rm dust} \approx 0.83 H_{\rm gas}$, respectively.

    \begin{figure}[t]
        \begin{center}
        \includegraphics[width=\hsize,bb = 0 0 443 671]{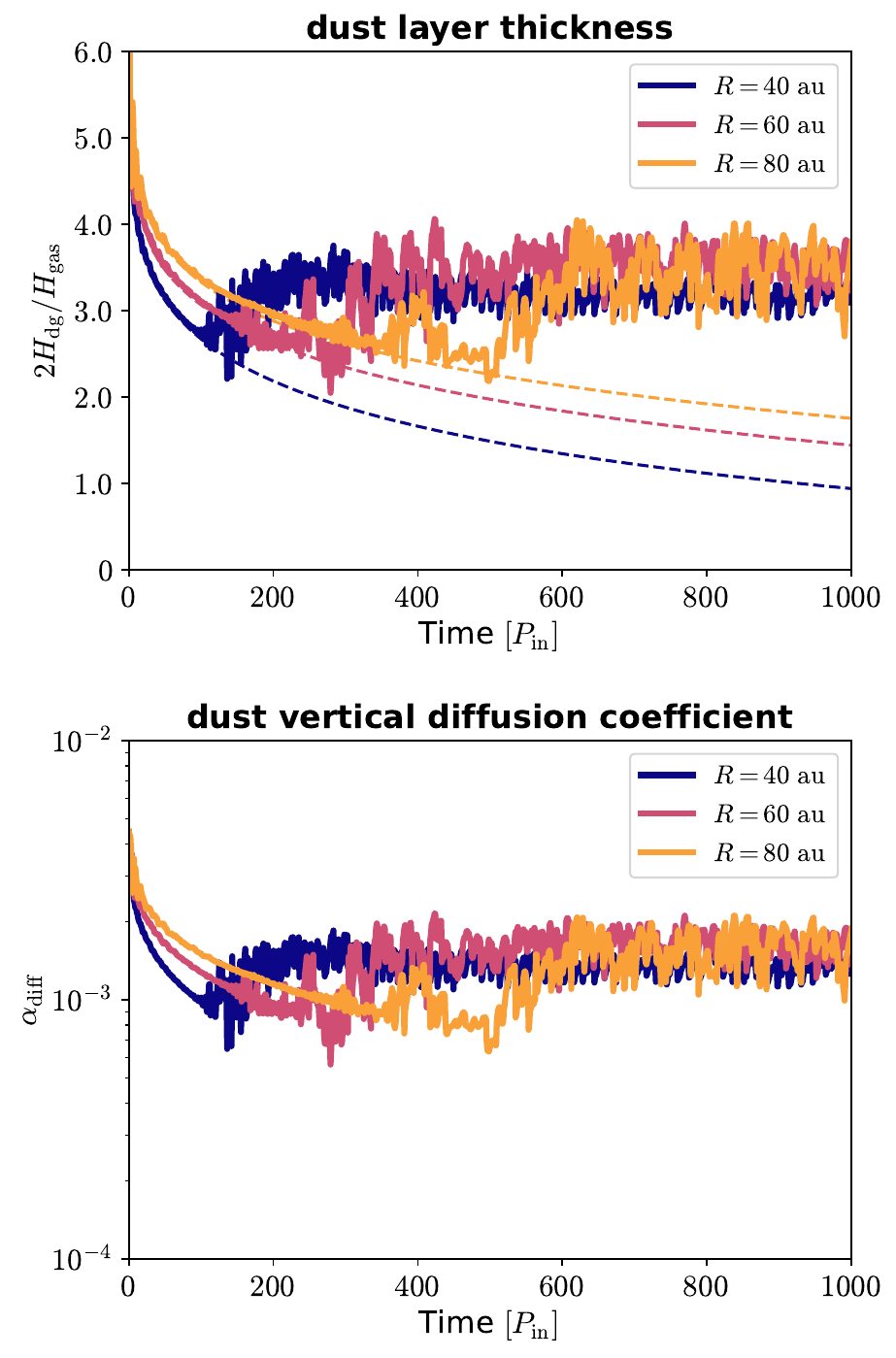}
        \end{center}
        \caption{Time evolution of dust layer thickness $2H_{\rm dg}/H_{\rm gas}$ (upper panel) and dimensionless dust vertical diffusion coefficient $\alpha_{\rm diff}$ (lower panel) for a run with $(Z_{\rm D2G},~a) = (0.01,~10~{\rm \mu m})$ in the dynamic-dust simulations with different values of $R$. The dashed lines in the upper panel show the grain's vertical position solved from Eq. \eqref{eq:dust_vertical_position} for each $R$.}
        \label{fig:2Hdg_alpha_TimeEvolution_D2G1e-2_a1e-3}
    \end{figure}  

    The thickness of the dust layer evolves with VSI-driven turbulence, as shown in Fig. \ref{fig:colormap_time_vzINI_rhodg_D2G1e-2_a1e-3_h}.
    The upper panel of Fig. \ref{fig:2Hdg_alpha_TimeEvolution_D2G1e-2_a1e-3} shows the evolution of $2H_{\rm dg}$ at $R = 40~{\rm au},~60~{\rm au}$, and $80~{\rm au}$ for the simulation displayed in Figs. \ref{fig:colormap_vz_rhodgINI_D2G1e-2_a1e-3_h} and \ref{fig:colormap_time_vzINI_rhodg_D2G1e-2_a1e-3_h}.
    During the early stage ($t \lesssim 200P_{\rm in}$), the dust layer becomes thinner as the dust settles toward the midplane.
    After this initial phase, the dust layer thickness reaches a quasi-steady value of $2H_{\rm dg} \approx 3$–$4H_{\rm gas}$.

    To assess whether VSI-driven turbulence can sustain the dust layer, we introduce a theoretical model of dust settling.
    The equation that describes the evolution of the grain's vertical position $z_{\rm dust}$ in the absence of gas turbulence is given by
    \begin{equation}\label{eq:dust_vertical_position}
        \frac{\mbox{d}^2z_{\rm dust}}{\mbox{d}t^2} = -\Omega_{\rm K}^2z_{\rm dust}-\frac{1}{t_{\rm stop}}\frac{\mbox{d}z_{\rm dust}}{\mbox{d}t}.
    \end{equation}
    This describes the dust vertical dynamics in the absence of gas motion.
    The dashed lines of Fig. \ref{fig:2Hdg_alpha_TimeEvolution_D2G1e-2_a1e-3} plot the theoretical prediction of dust settling using Eq. \eqref{eq:dust_vertical_position} with $z_{\rm dust} = \max{\{z\}}$ and $dz_{\rm dust}/dt = 0$ at $t = 0$.
    This comparison implies that dust settles undisturbed toward the midplane until approximately $150P_{\rm in}$ at $R = 40~{\rm au}$, after which turbulence inhibits further settling.

    Based on the thickness of the dust layer $H_{\rm dg}$, we estimate the dimensionless vertical diffusion coefficient $\alpha_{\rm diff}$ using Eq. \eqref{eq:H_dg}.
    The lower panel of Fig. \ref{fig:2Hdg_alpha_TimeEvolution_D2G1e-2_a1e-3} shows the evolution of $\alpha_{\rm diff}$ at $R = 40~{\rm au},~60~{\rm au}$, and $80~{\rm au}$ for the simulation illustrated in Figs. \ref{fig:colormap_vz_rhodgINI_D2G1e-2_a1e-3_h} and \ref{fig:colormap_time_vzINI_rhodg_D2G1e-2_a1e-3_h}.
    In the equilibrium state ($t \gtrsim 500P_{\rm in}$), $\alpha_{\rm diff}$ converges to approximately $2 \times 10^{-3}$, which is consistent with the typical value derived from the squared vertical velocity dispersion in VSI-driven turbulence (e.g., \citealt{FukuharaOkuzumi+:2023aa}).
    The correlation between these values allows us to measure the correlation timescale of VSI-driven turbulence (see appendix \ref{appendix:correlation_timescale}).

    \begin{figure*}[t]
        \begin{center}
        \includegraphics[width=\hsize,bb = 0 0 724 301]{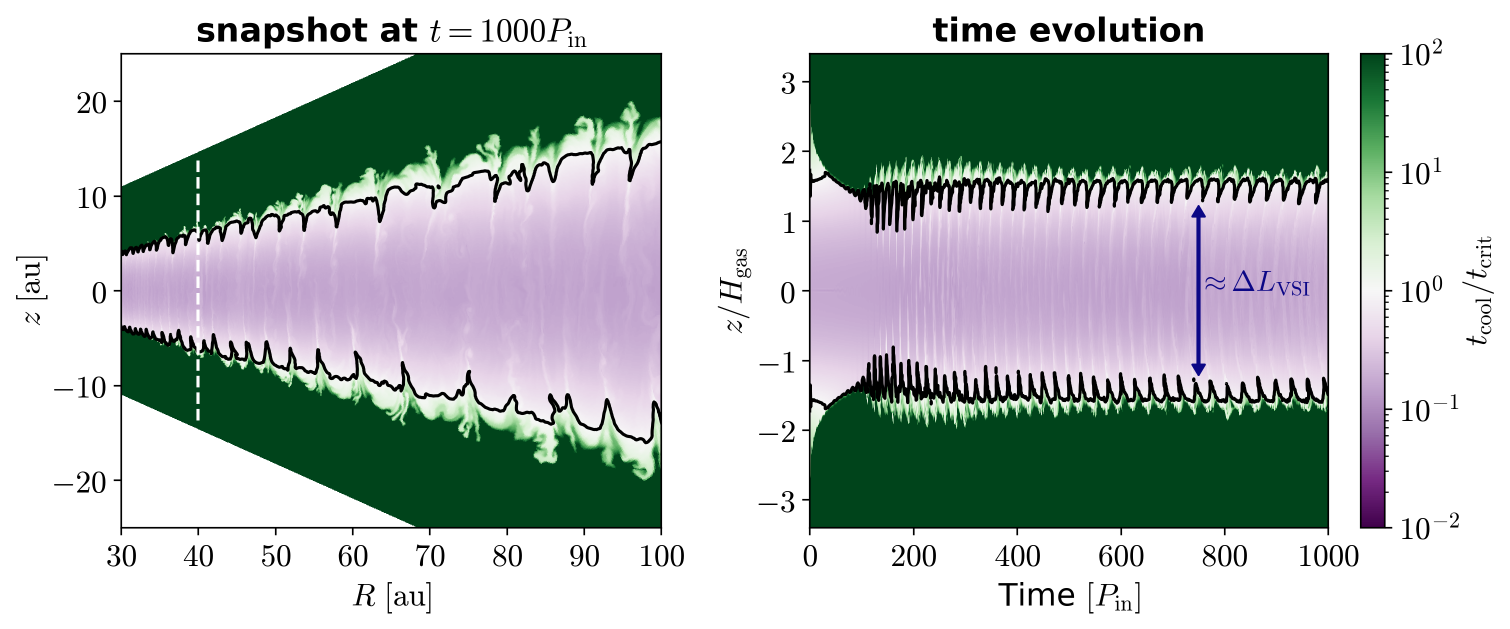}
        \end{center}
        \caption{Left panel: Cooling timescale $t_{\rm cool}$ normalized by the critical timescale $t_{\rm crit}$ for a run with $(Z_{\rm D2G},~a) = (0.01,~10~{\rm \mu m})$ in the dynamic-dust simulations, as a function of $R$ and $z$ at $t=1000P_{\rm in}$. The solid lines mark $t_{\rm cool}=t_{\rm crit}$. The dashed line indicates $R=40~{\rm au}$. Right panel: Time evolution of $t_{\rm cool}/t_{\rm crit}$ for the same run as the left panel at $R=40~{\rm au}$.}
        \label{fig:colormap_beta_D2G1e-2_a1e-3_h}
    \end{figure*}  

    In our simulations, the cooling timescale depends on the dust density with $t_{\rm cool}\propto \rho_{\rm dust}^{-1}$ [see Eqs. \eqref{eq:t_cool} and \eqref{eq:lgd}].
    Therefore, the VSI-active region maps a region nearly identical to the dust layer.
    We show in Fig. \ref{fig:colormap_beta_D2G1e-2_a1e-3_h} the snapshot (left panel) and time evolution (right panel) of the cooling timescale $t_{\rm cool}$ normalized by the critical timescale $t_{\rm crit}$.
    The region around the midplane corresponds to the VSI-active region, whose profile is similar to the dust layer's one shown in the right panels of Figs. \ref{fig:colormap_vz_rhodgINI_D2G1e-2_a1e-3_h} and \ref{fig:colormap_time_vzINI_rhodg_D2G1e-2_a1e-3_h}.

    \begin{figure}[t]
        \begin{center}
        \includegraphics[width=\hsize,bb = 0 0 453 320]{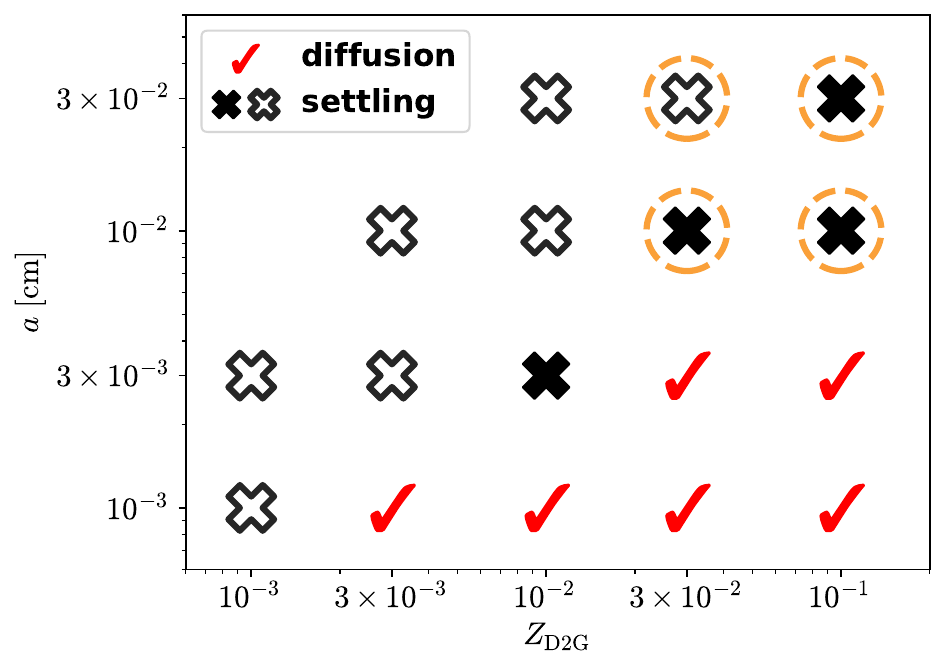}
        \end{center}
        \caption{Summary of the dynamic-dust simulations in terms of whether each simulation shows strong diffusion (check mark) or no diffusion (cross mark), mapped in the $Z_{\rm D2G}$--$a$ plane. Runs where the VSI-active region exists around the midplane in the initial condition are marked with filled crosses (and check mark), while runs with no VSI-active region in the initial condition are marked with empty crosses. The dashed circles represent runs for $\rho_{\rm dust}/\rho_{\rm gas}$ at the midplane that exceeds unity at the computational end time.}
        \label{fig:parameter_space_summary}
    \end{figure}  

    Because the cooling timescale is a function of the dust density $\rho_{\rm dust}$ and dust grain size $a$, one can expect that the condition to saturate to the equilibrium state where dust settling balances with diffusion depends on $Z_{\rm D2G}$ and $a$.
    To test this hypothesis, we perform simulations with different values of $Z_{\rm D2G}$ and $a$.
    Figure \ref{fig:parameter_space_summary} shows the summary of the simulations in terms of whether each run shows strong diffusion (check mark) or no diffusion (cross mark), mapped in the parameter space.
    When $a=10~{\rm \mu m}$ and $30~{\rm \mu m}$, the simulations with $Z_{\rm D2G}\geq 0.003$ and $\geq 0.03$ reach the equilibrium state where the turbulence diffuses the dust vertically.
    This is because the high dust density and small grain size lead to the short cooling timescale [Eqs. \eqref{eq:t_cool} and \eqref{eq:lgd}] and thereby the thick VSI-active layer around the midplane.

    At $Z_{\rm D2G}\leq 0.001$ for $a=10~{\rm \mu m}$ and $Z_{\rm D2G}\leq 0.01$ for $a=30~{\rm \mu m}$, the VSI-driven turbulence can not sustain the dust diffusion, and the dust settles toward the midplane, which is similar to large grains ($a\geq 100~{\rm \mu m}$).
    As an example of this case, Fig. \ref{fig:2Hdg_TimeEvolution_D2G1e-2_a1e-2} in appendix \ref{appendix:no_turbulence_case} shows the time evolution of the dust layer thickness for a run with $(Z_{\rm D2G},~a) = (0.01,~100~{\rm \mu m})$.

    In the settling cases, strong dust clumps can form due to the streaming instability (e.g., \citealt{YoudinGoodman:2005aa,LiYoudin:2021aa,LimSimon+:2024aa,LimSimon+:2025aa}).
    The dashed circles represent runs where $\rho_{\rm dust}/\rho_{\rm gas}$ at the midplane exceeds unity, which leads to the formation of dust clumps through the streaming instability (e.g., \citealt{LimSimon+:2024aa}).
    When $Z_{\rm D2G}\geq 0.03$ and $a\geq 100~{\rm \mu m}$, the dust may form strong clumps near the midplane because the dust settles in a runaway fashion due to the absence of VSI-driven turbulence.
    We note that our simulations do not resolve the streaming instability because of the absence of the aerodynamic feedback from dust to gas.

    \begin{figure}[t]
        \begin{center}
        \includegraphics[width=\hsize,bb = 0 0 434 327]{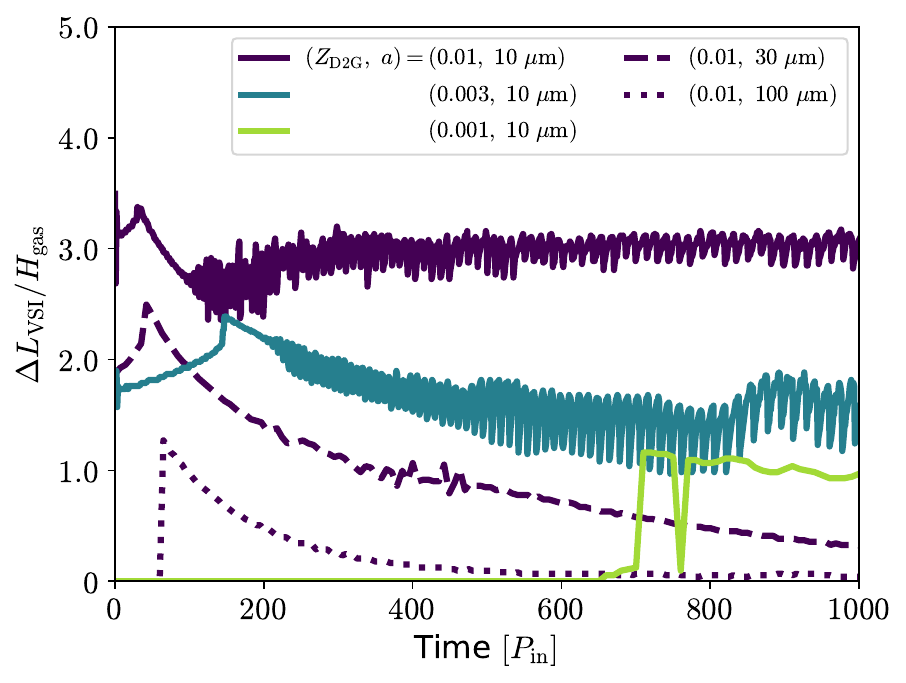}
        \end{center}
        \caption{Time evolution of the VSI-active region's thickness $\Delta L_{\rm VSI}/H_{\rm gas}$ for runs with different values of the dust-to-gas mass ratio $Z_{\rm D2G}$ and dust grain size $a$ in the dynamic-dust simulations at $R=40~{\rm au}$.}
        \label{fig:VSIregion_thickness}
    \end{figure} 

    Even if no VSI-active region is present in the initial condition (empty crosses of Fig. \ref{fig:parameter_space_summary}), it forms as the dust settles.
    We plot in Fig. \ref{fig:VSIregion_thickness} the evolution of the VSI-active region's thicknesses for runs with different values of $Z_{\rm D2G}$ and $a$.
    For the runs with $(Z_{\rm D2G},~a) = (0.001,~10~{\rm \mu m})$ and $(0.01,~100~{\rm \mu m})$, the VSI-active regions, which are not initially present, form at $t\approx 70P_{\rm in}$ and $700P_{\rm in}$ with thicknesses of $\Delta L_{\rm VSI}\approx1.3H_{\rm gas}$ and $1.5H_{\rm gas}$, respectively. 
    However, these VSI-active regions are too thin to generate strong turbulence, and the regions shrink thereafter with dust settling.
    This is because the intensity of VSI-driven turbulence decreases sharply as $\Delta L_{\rm VSI}$ falls below $2H_{\rm gas}$ (see also \citealt{FukuharaOkuzumi+:2023aa}).
    Even if the VSI-active regions exist initially, the final state is also different.
    The runs with $(Z_{\rm D2G},~a) = (0.003,~10~{\rm \mu m})$ and $(0.01,~30~{\rm \mu m})$ have the same vertical distribution of the cooling time and therefore the same thickness of the VSI-active region at the initial condition.
    In the small dust case, the VSI can drive turbulence because its intensity is enough to sustain the thick dust and VSI-active layers.
    On the other hand, in the large dust case, the dust layer is too thin to maintain the thick VSI-active layer needed to drive strong turbulence.
    This is because the thickness of the VSI-active layers, as well as of the dust layers, depends on the turbulence intensity and grain size, and also the intensity saturates at a certain level.
    In addition to this factor, the suppression of turbulence by large dust particles can be due to large grains settling quickly before turbulence develops, making the VSI-active layer thin.

    \begin{figure}[t]
        \begin{center}
        \includegraphics[width=\hsize,bb = 0 0 432 331]{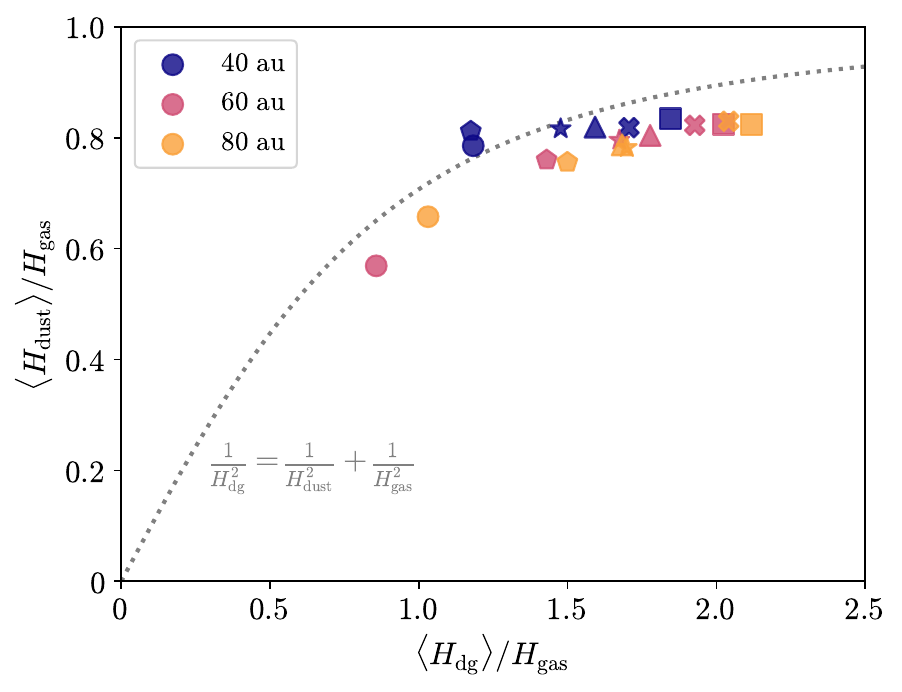}
        \end{center}
        \caption{Time-average half-thicknesses of the dust layer $\langle H_{\rm dg}\rangle$ vs. time-average dust scale height $\langle H_{\rm dust}\rangle$ with different values of $R$ for runs of diffusion cases in the dynamic-dust simulations. The symbols correspond to runs with $(Z_{\rm D2G},~a) = (0.003,~10~{\rm \mu m})$ (circles), $(0.01,~10~{\rm \mu m})$ (triangles), $(0.03,~10~{\rm \mu m})$ (crosses), $(0.1,~10~{\rm \mu m})$ (squares), $(0.03,~30~{\rm \mu m})$ (pentagons), and $(0.1,~30~{\rm \mu m})$ (stars). The dotted line marks the relationship of $H_{\rm dg}$ and $H_{\rm dust}$ following Eq. \eqref{eq:relation_Hdg_Hd}.}
        \label{fig:Hdg_vs_Hd}
    \end{figure}  

    In the diffusion cases, the thicknesses of the dust layer $2H_{\rm dg}$ range from $2H_{\rm gas}$ to $4H_{\rm gas}$, corresponding to dust scale heights $H_{\rm dust}$ of $\approx 0.6$--$0.8 H_{\rm gas}$.
    Figure \ref{fig:Hdg_vs_Hd} plots the time-average half-thicknesses of the dust layer $\langle H_{\rm dg}\rangle$ versus time-average dust scale height $\langle H_{\rm dust}\rangle$ for runs where dust vertical profiles reach the equilibrium diffusion states.
    The dotted curve marks the correlation between $H_{\rm dg}$ and $H_{\rm dust}$ given by Eq. \eqref{eq:relation_Hdg_Hd}.
    This figure indicates that the relationship between the dust layer thicknesses and dust scale height in VSI-driven turbulence approximately follows that for the turbulent vertical mixing-settling balance.

    \subsection{Equilibrium state with fixed-to-dynamic-dust simulations}\label{subsec:equilibrium_state_turbulent}

    The dynamic-dust simulations presented in the previous subsection show that VSI-driven turbulence operates and diffuses dust only when $Z_{\rm D2G}\geq 0.003$ and $\geq 0.03$ for $a=10~{\rm \mu m}$ and $30~{\rm \mu m}$, respectively.
    We find that these requirements hold no matter whether or not VSI-driven turbulence operates initially.

    \begin{figure}[t]
        \begin{center}
        \includegraphics[width=\hsize,bb = 0 0 453 320]{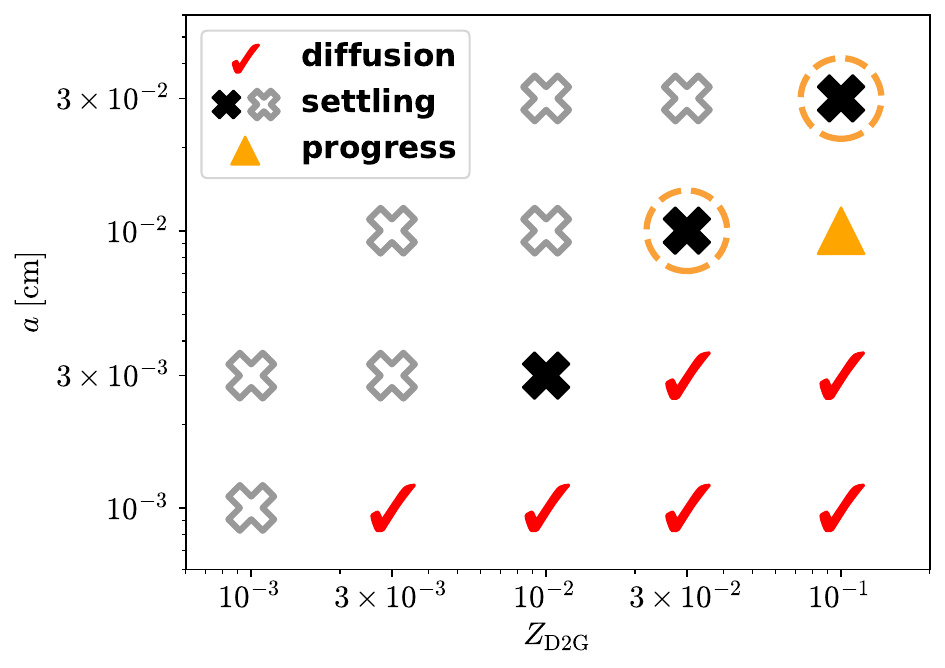}
        \end{center}
        \caption{Summary of the fixed-to-dynamic-dust simulations in terms of whether each simulation shows strong diffusion (check mark) or no diffusion (cross mark), mapped in the $Z_{\rm D2G}$--$a$ plane. The empty crosses plot the runs that have not been calculated for the fixed-to-dynamic-dust simulations because there is no VSI-active region initially. The triangle marks plot the runs where evolution is in progress. The dashed circles represent runs for $\rho_{\rm dust}/\rho_{\rm gas}$ at the midplane that exceeds unity at the computational end time.}
        \label{fig:parameter_space_summary_turb}
    \end{figure}

    Figure \ref{fig:parameter_space_summary_turb} summarizes the fixed-to-dynamic-dust simulations, indicating whether each run exhibits strong diffusion (check mark) or settling (cross mark) across the parameter space.
    The runs with $Z_{\rm D2G}$ and $a$ that reach a strong-diffusion state in the dynamic-dust simulations also achieve the same state in the fixed-to-dynamic-dust simulations.
    The run with $(Z_{\rm D2G},~a) = (0.1,~100~{\rm \mu m})$ is for a state in which equilibrium has not been reached and turbulence is not completely eliminated in the simulated time.
    In this case, VSI-driven turbulence gradually decays over roughly $1000P_{\rm in}$, and it is expected that the system will eventually evolve toward a state where turbulence is suppressed in the entire region and the dust fully settles on the midplane.
    Figures \ref{fig:colormap_time_vz_rhodgINI_turb_D2G1e-1_a1e-2_h_1500} and \ref{fig:colormap_vz_D2G1e-1_a1e-2_turb_time} in appendix \ref{appendix:progress_case} illustrate the time evolution of $v_{z,\rm gas}$ and $\rho_{\rm dust}/\rho_{\rm gas}$ at $R = 40~{\rm au}$ and snapshots of $v_{z,\rm gas}$ in the $R$--$z$ plane at different times, respectively.
    Based on these results, we conclude that runs with the same values of $Z_{\rm D2G}$ and $a$ reach comparable states in both the dynamic-dust and fixed-to-dynamic-dust simulations.
    For runs with the large dust (cross mark), VSI-driven turbulence is rapidly suppressed, reaching a state where dust continues to settle on the midplane.

    \begin{figure*}[t]
        \begin{center}
        \includegraphics[width=\hsize,bb = 0 0 826 272]{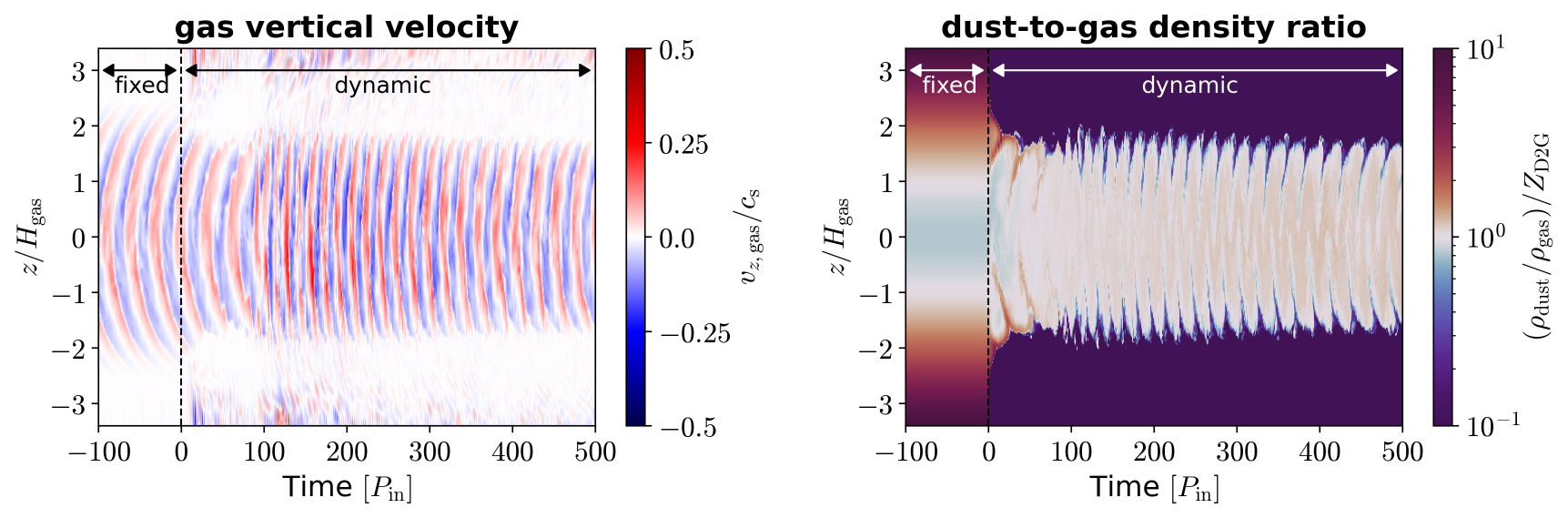}
        \end{center}
        \caption{Gas vertical velocity $v_{z,\rm gas}/c_{\rm s}$ (left) and ratio of dust and gas densities $(\rho_{\rm dust}/\rho_{\rm gas})/Z_{\rm D2G}$ (right) for a run with $(Z_{\rm D2G},~a) = (0.01,~10~{\rm \mu m})$ in the fixed-to-dynamic-dust simulations, as a function of time and $z/H_{\rm gas}$ at $R=40~{\rm au}$. We fix the dust and cooling timescale until $t=0$ and switch to dynamic dust after that.}
        \label{fig:colormap_time_vz_rhodgINI_turb_D2G1e-2_a1e-3_h}
    \end{figure*} 

    To see how it transitions to the equilibrium state, we plot in Fig. \ref{fig:colormap_time_vz_rhodgINI_turb_D2G1e-2_a1e-3_h} the time evolution of the vertical profile of $v_{z,\rm gas}$ and $\rho_{\rm dust}/\rho_{\rm gas}$ for a run with $(Z_{\rm D2G},~a) = (0.01,~10~{\rm \mu m})$ at $R=40~{\rm au}$.
    Until $t=0$, the spatial profile of the cooling time is fixed to the initial conditions; thereafter, the dynamic cooling time and dust are enabled.
    Immediately afterward, the thickness of regions where VSI-driven turbulence operates is also narrowed as the dust settles toward the midplane.
    Subsequently, the turbulence and dust reach the equilibrium state where settling and diffusion are balanced, as shown in Sect. \ref{subsec:equilibrium_state_laminar}.

    \begin{figure}[t]
        \begin{center}
        \includegraphics[width=\hsize,bb = 0 0 451 327]{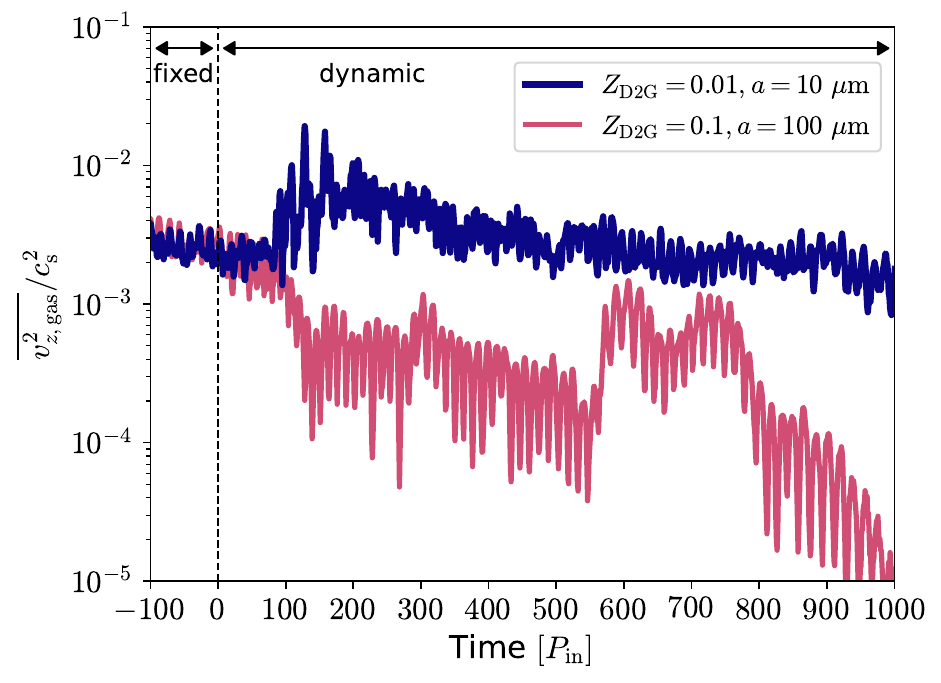}
        \end{center}
        \caption{Time evolution of vertical mean squared gas velocity $\overline{v_{z,\rm gas}^2}/c_{\rm s}^2$ for runs with $(Z_{\rm D2G},~a) = (0.01,~10~{\rm \mu m})$ and $(0.1,~100~{\rm \mu m})$ in the fixed-to-dynamic-dust simulations, at $R=40~{\rm au}$.}
        \label{fig:squared_vertical_velocity_turb}
    \end{figure}

    To quantify the evolution of turbulence strength, Fig. \ref{fig:squared_vertical_velocity_turb} shows the time evolution of the squared vertical gas velocity averaged over the vertical direction, $\overline{v_{z,\rm gas}^2}$, for the two cases presented in Figs. \ref{fig:colormap_time_vz_rhodgINI_turb_D2G1e-2_a1e-3_h} and \ref{fig:colormap_time_vz_rhodgINI_turb_D2G1e-1_a1e-2_h_1500}.
    For the strong diffusion case with $(Z_{\rm D2G},~a) = (0.01,~10~{\rm \mu m})$, $\overline{v_{z,\rm gas}^2}$ remains at a level of approximately $2$--$5 \times 10^{-3} c_{\rm s}^2$, even after switching from fixed to dynamic dust calculations.
    In contrast, for the run with $(Z_{\rm D2G},~a) = (0.1,~100~{\rm \mu m})$, $\overline{v_{z,\rm gas}^2}$ drops to well below $10^{-5} c_{\rm s}^2$ within about $1000 P_{\rm in}$ after the switch to dynamic dust.

    \begin{figure}[t]
        \begin{center}
        \includegraphics[width=\hsize,bb = 0 0 443 671]{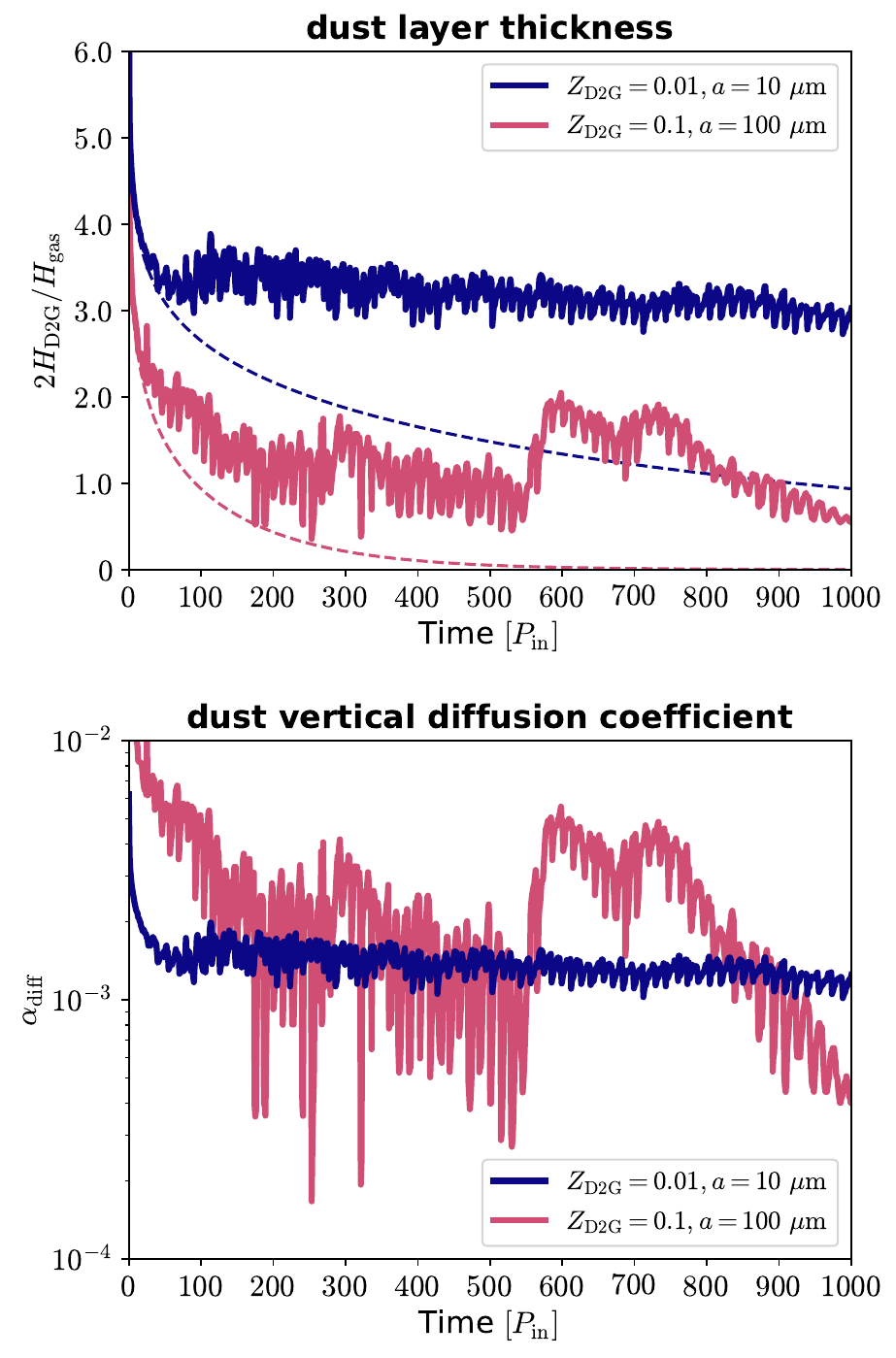}
        \end{center}
        \caption{Time evolution of dust layer thickness $2H_{\rm dg}/H_{\rm gas}$ (upper panel) and dimensionless dust vertical diffusion coefficient $\alpha_{\rm diff}$ (lower panel) for runs with $(Z_{\rm D2G},~a) = (0.01,~10~{\rm \mu m})$ and $(0.1,~100~{\rm \mu m})$ in the fixed-to-dynamic-dust simulations, at $R=40~{\rm au}$. The dashed lines show the grain's vertical position solved from Eq. \eqref{eq:dust_vertical_position}.}
        \label{fig:2Hdg_alpha_TimeEvolution_turb}
    \end{figure}

    For the fixed-to-dynamic-dust simulations, the evolution of the dust layer thickness is different from the dynamic-dust simulations in Sect. \ref{subsec:equilibrium_state_laminar} because the dust settles with turbulence already driven.
    In Fig. \ref{fig:2Hdg_alpha_TimeEvolution_turb}, we show the time evolution of the dust layer thickness (upper panel) and dust vertical diffusion coefficient (lower panel) for runs displayed in Figs. \ref{fig:colormap_time_vz_rhodgINI_turb_D2G1e-2_a1e-3_h} and \ref{fig:colormap_time_vz_rhodgINI_turb_D2G1e-1_a1e-2_h_1500}.
    In the case of $(Z_{\rm D2G},~a) = (0.01,~10~{\rm \mu m})$, the dust layer quickly relaxes an equilibrium state with a thickness of $\approx 3H_{\rm gas}$, leading to $\alpha_{\rm diff}\approx 10^{-3}$.
    This equilibrium state is almost the same as in the case of the dynamic-dust simulation (see Fig. \ref{fig:2Hdg_alpha_TimeEvolution_D2G1e-2_a1e-3}).
    In the case of $(Z_{\rm D2G},~a) = (0.1,~100~{\rm \mu m})$, the dust eventually settles on the midplane, although it is disturbed by turbulence.
    The discrepancy from the theoretical prediction (dashed line) suggests the presence of a positive feedback process.
    In this process, the VSI-active region shrinks by dust settling even though the VSI drives turbulence, leading to weaker turbulence intensity and further promoting dust settling.
    This mechanism is equivalent to the runaway settling described by \citet{FukuharaOkuzumi:2024aa}.
    
\section{Discussion}\label{sec:discussion}

    \subsection{What determines the existence of equilibrium state?}\label{subsec:existence_equilibrium}

    Our results suggest that the initial vertical distribution of the cooling timescale does not determine the existence of an equilibrium state between VSI-driven turbulence and dust.
    Larger dust sizes and dust-to-gas mass ratios do not allow VSI-driven turbulence to operate, even with the same initial cooling rate distribution (see Figs. \ref{fig:parameter_space_summary} and \ref{fig:parameter_space_summary_turb}).

    We expect this condition to depend on the feasibility of balancing dust settling and turbulent diffusion in equilibrium states.
    Our results in this study indicate that VSI-driven turbulence only reaches a vertical diffusion coefficient of at most $\approx 2\times 10^{-3}$ when driven.
    As expected in \citet{FukuharaOkuzumi:2024aa}, VSI-driven turbulence may maintain the dust distribution if the thickness of the VSI-active region created by the dust under this diffusion exceeds two gas scale heights.

    To test this hypothesis, we calculate the equilibrium vertical diffusion coefficient $\alpha_{z,\rm equi}$ for different values of the dust-to-gas mass ratio $Z_{\rm D2G}$ and dust grain size $a$, using the semi-analytic model of \citet{FukuharaOkuzumi:2024aa}.
    This is the self-consistent model that determines the vertical dust distribution and VSI-driven turbulence intensity.
    We apply the model for the single-sized dust to the disk of this study and obtain the prediction of the equilibrium state where dust settling balances with VSI-driven turbulent diffusion.

    \begin{figure}[t]
        \begin{center}
        \includegraphics[width=\hsize,bb = 0 0 441 318]{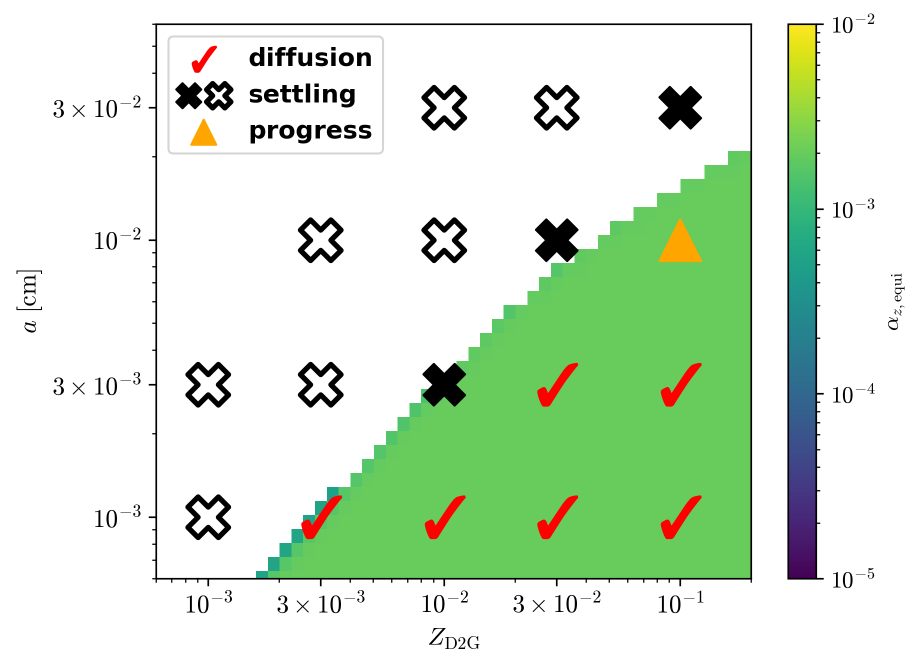}
        \end{center}
        \caption{Vertical diffusion coefficient $\alpha_{z,\rm equi}$ for the equilibrium solution of the model shown in \citet{FukuharaOkuzumi:2024aa}, as a function of the dust-to-gas mass ratio $Z_{\rm D2G}$ and dust grain size $a$. The white area indicates the parameter space where no equilibrium solution exists. The check, cross, and triangle marks show the summary of the fixed-to-dynamic-dust simulations, as shown in Fig. \ref{fig:parameter_space_summary_turb}.}
        \label{fig:prediction_FO24}
    \end{figure}

    Figure \ref{fig:prediction_FO24} shows $\alpha_{z,\rm equi}$ as a function of $Z_{\rm D2G}$ and $a$.
    This figure indicates that the model of \citet{FukuharaOkuzumi:2024aa} can predict the equilibrium state for dust and VSI-driven turbulence.
    Therefore, we conclude that the balance between dust vertical turbulent diffusion and settling determines whether equilibrium states for VSI-driven turbulence and dust vertical profile can be achieved.
    For the runs with $(Z_{\rm D2G},~a) = (0.01,~30~{\rm \mu m})$ and $(0.1,~100~{\rm \mu m})$, however, the dust settles toward the midplane due to the absence of turbulence, contrary to expectations of the analytic model.
    This difference may be because the vertical dust profiles of our simulation deviate slightly from the Gaussian profile, which is assumed in the semi-analytic model.

    \subsection{Dust growth in VSI-turbulent disk}\label{subsec:dust_growth}

    We find that the vertical dust distribution in a VSI-dominated disk depends on both dust grain size and total dust amount.
    For small grains, VSI-driven turbulence sustains strong vertical diffusion with $\alpha_{\rm diff} \sim 10^{-3}$, while large grains lead to significant settling due to the absence of such turbulence.
    These results suggest that micron-sized grains initially grow under strong VSI-driven turbulence into (sub)millimeter sizes, after which further growth occurs in a largely laminar environment.
    This transition may promote the formation of kilometer-sized planetesimals via the streaming instability (e.g., \citealt{YoudinGoodman:2005aa,LiYoudin:2021aa,LimSimon+:2024aa,LimSimon+:2025aa}) and secular gravitational instability (e.g., \citealt{Ward:2000aa,Youdin:2011aa,TominagaInutsuka+:2023aa}).
    This dust growth process in VSI-dominated disks has the potential for efficient planetesimal formation.
    We plan to explore this possibility in future work.

    The VSI-active region is confined not only vertically but also radially within the disk.
    In regions where VSI-driven turbulence operates, dust can move inward faster than by radial drift (see Fig. \ref{fig:dust_radial_velocity}).
    At the inner disk region that is optically thick to dust thermal emission, the cooling would be inefficient and thus suppress the VSI, creating the inner edge of the VSI-active region \citep{FukuharaOkuzumi:2024aa}.
    At this inner radial boundary, dust can accumulate, as the radial dust velocity inside the VSI-active region is regulated primarily by radial drift.
    This is a potential candidate location where planetesimals could form efficiently.
    To evaluate these, hydrodynamical simulations incorporating dust dynamics near the VSI-active region's edge should be performed.

    \subsection{Limitations of our simulations}\label{subsec:Limitation}

    Our simulations have several important limitations that should be addressed in future work.
    First, we consider only the collisional timescale as the cooling timescale.
    In reality, the cooling rate depends on the optical depth (e.g., \citealt{Malygin+2017}).
    In optically thick inner disk regions, radiative diffusion dominates the cooling process, making cooling inefficient (e.g., \citealt{FukuharaOkuzumi:2024aa}).
    This suggests the presence of an inner edge to the VSI-active region.
    A steep radial gradient in turbulent viscosity at such an edge could trigger the Rossby wave instability \citep{LovelaceLi+:1999sg,LiFinn+:2000aa,LiColgate+:2001gv}, potentially forming long-lived vortices that concentrate dust and facilitate planetesimal formation via gravitational collapse (e.g., \citealt{BargeSommeria:1995qd}).
    However, it is unclear whether this edge or vortex can be sustained in dust-controlled VSI-driven turbulence because changes in the dust distribution affect the cooling time distribution.
    Understanding the role of the VSI-active region's edge in planetesimal formation requires three-dimensional hydrodynamical simulations with dynamically evolving cooling rates that respond to radial variations in the dust distribution.

    Our simulations also assume only single-sized dust particles.
    Considering the dust size distribution, the dust size that can achieve an equilibrium state between dust and VSI-driven turbulence can vary \citep{FukuharaOkuzumi:2024aa}.
    Our condition for the equilibrium state would be a condition of effective dust size that dominates the cooling rate.
    Recently, \citet{PfeilBirnstiel+:2024aa} have partially considered the dust growth and size distribution to simulate VSI-driven turbulence in dynamic cooling rate.
    They suggest that the onset of turbulent gas flows requires a large initial dust scale height, typically resulting from small dust grains.
    In contrast, our results suggest that, unless the initial dust scale height is extremely small, the existence of the equilibrium state is governed not by it, but by the balance between turbulent diffusion and dust settling in the saturated state (see Sect. \ref{subsec:existence_equilibrium}).
    These findings lead us to expect that, even with multiple dust species of different sizes, the balance between dust settling and turbulent diffusion still primarily regulates the dust vertical distribution and the VSI-driven turbulence structure, as with the single-sized dust\footnote{\citet{FukuharaOkuzumi:2024aa} investigated the equilibrium state of the dust vertical distribution and VSI-driven turbulence for both dust grains with a size distribution and single-sized dust grains, using the semi-analytic model presented in their study.}.
    When small grains still dominate cooling, an equilibrium state can exist even if the maximum dust grains are millimeter-sized.
    Therefore, a requirement of the VSI on dust size distribution should be evaluated by accounting for both turbulent diffusion and thermal coupling for differently sized grains.
    Specific dust growth processes in non-isotropic turbulence such as VSI-driven turbulence have also not been well investigated.
    It is necessary to investigate at what point in the dust growth the VSI-driven turbulence dies in the future.

    Furthermore, this study ignores the effects of dust loading, magnetic field, and vertical thermal structure.
    Dust would increase the effective gas buoyancy frequency that prevents the growth of the linear VSI \citep{Lin:2017aa}.
    The gas--dust drag force can also suppress the VSI and dust diffusion \citep{Lin:2019aa,LehmannLin:2022nr,LehmannLin:2023aa}, and may enable the coexistence of the VSI and streaming instability (e.g., \citealt{SchaferJohansen+:2025aa,HuangBai:2025aa}).
    We note that, however, the effect of dust loading would play only a minor role on the equilibrium state in this study because of $\rho_{\rm dust}/\rho_{\rm gas} \ll 1$ and $\partial_z(\rho_{\rm dust}/\rho_{\rm gas}) \approx 0$ within the dust layers (\citealt{Lin:2019aa}; see Figs. \ref{fig:vertical_profile_rhod} and \ref{fig:Z_saturated_mid})\footnote{The dust-induced vertical buoyancy frequency $N_{z,\rm d}$, which characterizes the effect of dust on the vertical buoyancy stabilizing vertical motion of the VSI, depends on the ratio of the gas and dust densities and its vertical gradient, given by $N_{z,\rm d}^2 \propto (1+\rho_{\rm dust}/\rho_{\rm gas})^{-1}\partial_z(\rho_{\rm dust}/\rho_{\rm gas})$ [see Eq. (17) of \citealt{Lin:2019aa}].}.
    Moreover, magnetic fields threading the global disk may suppress the VSI either directly through magnetic tension or indirectly through MRI-driven turbulence \citep{NelsonGresselUmurhan2013,LatterPapaloizou2018,Cui:2020aa}.
    The roles of magnetic fields in the VSI suppression can depend on non-ideal magnetohydrodynamical effects (ambipolar diffusion, Ohmic resistivity, and Hall effect; \citealt{Cui:2020aa,CuiBai:2022aa,CuiLin:2021cj,LatterKunz:2022ic}).
    Additionally, a vertical thermal stratification with a colder interior and a hotter surface can change the structure and intensity of VSI-driven turbulence at the region around the midplane \citep{ZhangZhu+:2024aa,YunKim+:2025aa,YunKim+:2025ab}.
    These effects would change the criterion of the equilibrium state between the VSI and dust, thereby they should be considered in future work.

\section{Conclusions}\label{sec:conclusions}
We have investigated how the VSI drives turbulence with a dynamic cooling timescale and how it maintains dust diffusion. 
The local cooling is determined fully dynamically from the local dust density at each cell and timestep.
We have performed global 2.5-dimensional hydrodynamical simulations, including the dust modeled as a pressureless fluid and local $\beta$ cooling rates coupled to the dust density in an axisymmetric disk.
There are two types of simulations: (i) calculations with dust distribution and cooling timescale evolved from the beginning of the simulation, which we call dynamic-dust simulations,  and (ii) calculations with cooling timescale fixed in time and space until turbulence saturates to a quasi-steady state, followed by dynamic dust and cooling timescale activation, which we call fixed-to-dynamic-dust simulations. 
Our key findings are summarized as follows.
\begin{enumerate}
    \item From the dynamic-dust simulations, we find that for the small dust size, there exists the equilibrium state where VSI-driven turbulent diffusion balances with dust settling (Figs. \ref{fig:colormap_vz_rhodgINI_D2G1e-2_a1e-3_h} and \ref{fig:colormap_time_vzINI_rhodg_D2G1e-2_a1e-3_h}). In the equilibrium state, turbulence sustains the dust layer with a thickness of $\approx 3H_{\rm gas}$ and an associated VSI-active region of similar thickness around the midplane (Figs. \ref{fig:vertical_profile_rhod} and \ref{fig:colormap_beta_D2G1e-2_a1e-3_h}). The dimensionless dust vertical diffusion coefficient estimated by Eq. \eqref{eq:H_dg} reaches $\approx2\times 10^{-3}$ (Fig. \ref{fig:2Hdg_alpha_TimeEvolution_D2G1e-2_a1e-3}).
    \item The ability of the VSI to maintain the equilibrium state for dust settling and turbulent diffusion depends on the dust grain size $a$ and dust-to-gas mass ratio $Z_{\rm D2G}$ (Fig. \ref{fig:parameter_space_summary}). In the cases of large grains, the dust settles toward the midplane due to the absence of turbulence. The VSI-active regions exist for all runs presented in this study, either from the beginning or in the middle of the calculation. However, runs of large dust size and low dust-to-gas mass ratio lead to a thin VSI-active region and absence of turbulence (Fig. \ref{fig:VSIregion_thickness}).
    \item We find that in both dynamic-dust and fixed-to-dynamic-dust simulations, runs with the same values of $a$ and $Z_{\rm D2G}$ reach comparable states, regardless of whether turbulence is present initially or not (Fig. \ref{fig:parameter_space_summary_turb}). The same equilibrium state exists for small grains and high dust-to-gas mass ratio in the fixed-to-dynamic-dust simulations as in dynamic-dust simulations (Fig. \ref{fig:colormap_time_vz_rhodgINI_turb_D2G1e-2_a1e-3_h}). The value of the vertical diffusion coefficient is $\approx 2\times 10^{-3}$, which is almost the same as in the dynamic-dust simulations (Fig. \ref{fig:2Hdg_alpha_TimeEvolution_turb}).
    \item The dependence of the VSI-driven turbulence onset on $a$ and $Z_{\rm D2G}$ is consistent with the prediction of the semi-analytic model that determines the vertical dust distribution and the strength of VSI-driven turbulence in a self-consistent manner (\citealt{FukuharaOkuzumi:2024aa}; Fig. \ref{fig:prediction_FO24}).
\end{enumerate}

Our results suggest that in the VSI-dominated disk, dust grows under turbulence of different intensities for different dust sizes.
Under this environment, the efficiency of planetesimal formation would vary with the distribution of the VSI-active regions.
In order to quantify this effect, the detailed process of dust growth in VSI-driven turbulence should be investigated.

\begin{acknowledgements}
    We thank Takahiro Ueda and David Melon Fuksman for the useful discussion.
    We also appreciate the anonymous referee for comments that greatly helped improve the manuscript.
    Numerical computations were carried out on Cray XC50 and XD2000 at Center for Computational Astrophysics, National Astronomical Observatory of Japan, and the supercomputer in Max--Planck Institute for Astronomy.
    This work was supported by the National Science and Technology Council (grant 113-2112-M-001-036-) and an Academia Sinica Career Development Award (grant AS-CDA-110-M06). 
    This work was also supported by JSPS KAKENHI Grant Numbers JP22KJ1337, JP23K25923, and JP24KJ1041, and the JSPS Overseas Challenge Program for Young Researchers (202380168).
    M.F. acknowledge support from the European Research Council (ERC), under the European Union’s Horizon 2020 research and innovation program (grant agreement No. 757957).
\end{acknowledgements}

\bibliographystyle{aa}
\bibliography{FukuharaFlock25}

\begin{appendix}

\section{Dust concentration at the midplane}\label{appendix:dust_consentration}

\begin{figure}[t]
    \begin{center}
    \includegraphics[width=\hsize,bb = 0 0 445 328]{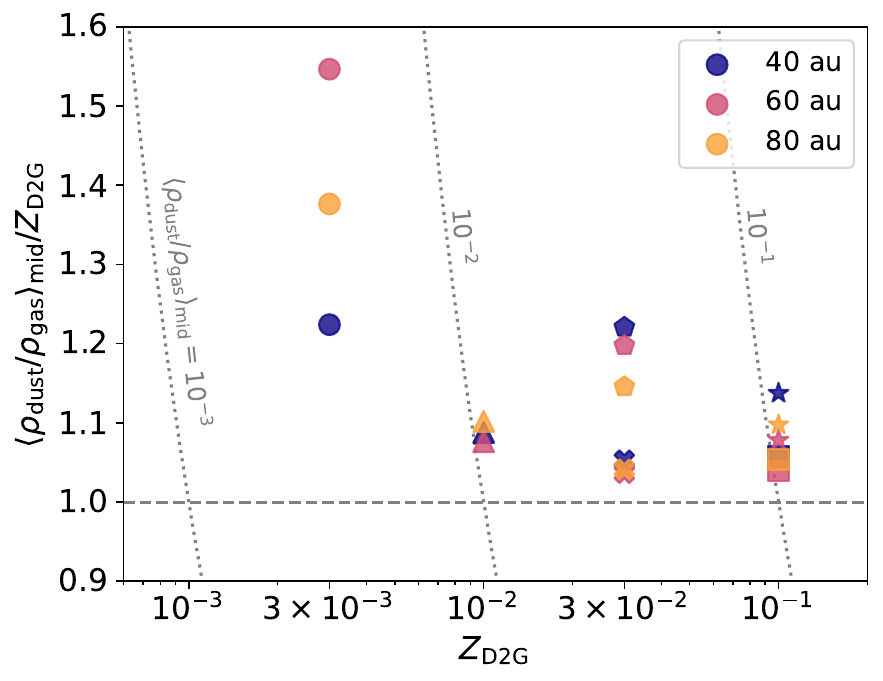}
    \end{center}
    \caption{Time-average ratio of dust and gas densities at the midplane $\langle\rho_{\rm dust}/\rho_{\rm gas}\rangle_{\rm mid}$ normalized by the initial dust-to-gas mass ratio $Z_{\rm D2G}$, with different values of $R$ for runs of diffusion cases in dynamic-dust simulations. The symbols correspond to runs with $(Z_{\rm D2G},~a) = (0.003,~10~{\rm \mu m})$ (circles), $(0.01,~10~{\rm \mu m})$ (triangles), $(0.03,~10~{\rm \mu m})$ (crosses), $(0.1,~10~{\rm \mu m})$ (squares), $(0.03,~30~{\rm \mu m})$ (pentagons), and $(0.1,~30~{\rm \mu m})$ (stars). The horizontal dashed line represents $\langle\rho_{\rm dust}/\rho_{\rm gas}\rangle_{\rm mid} = Z_{\rm D2G}$. The dotted lines show $\langle\rho_{\rm dust}/\rho_{\rm gas}\rangle_{\rm mid} = 10^{-3}$, $10^{-2}$, and $10^{-1}$.}
    \label{fig:Z_saturated_mid}
\end{figure}  

Figures \ref{fig:colormap_vz_rhodgINI_D2G1e-2_a1e-3_h}, \ref{fig:colormap_time_vzINI_rhodg_D2G1e-2_a1e-3_h}, and \ref{fig:vertical_profile_rhod} in Sect. \ref{subsec:equilibrium_state_laminar} show that the dust becomes slightly concentrated within the dust layers.
In Fig. \ref{fig:Z_saturated_mid}, we present the time-averaged dust-to-gas density ratio at the midplane, $\langle\rho_{\rm dust}/\rho_{\rm gas}\rangle_{\rm mid}$, for the diffusion runs in the dynamic-dust simulations.
This figure shows that, in most cases, the midplane dust-to-gas density ratio increases by a factor of approximately 1.05–1.2 compared to the initial value.
For a fixed dust grain size, $\langle\rho_{\rm dust}/\rho_{\rm gas}\rangle_{\rm mid}/Z_{\rm D2G}$ tends to decrease as $Z_{\rm D2G}$ increases.
Conversely, for a fixed $Z_{\rm D2G}$, $\langle\rho_{\rm dust}/\rho_{\rm gas}\rangle_{\rm mid}/Z_{\rm D2G}$ tends to increase as the dust grain size becomes larger.
These trends are tentative and suggest that the degree of dust concentration may depend on both the initial dust-to-gas mass ratio and grain size.

\section{Dust radial velocity in VSI-driven turbulence}\label{appendix:dust_radial_velocity}

\begin{figure}[t]
    \begin{center}
    \includegraphics[width=\hsize,bb = 0 0 445 328]{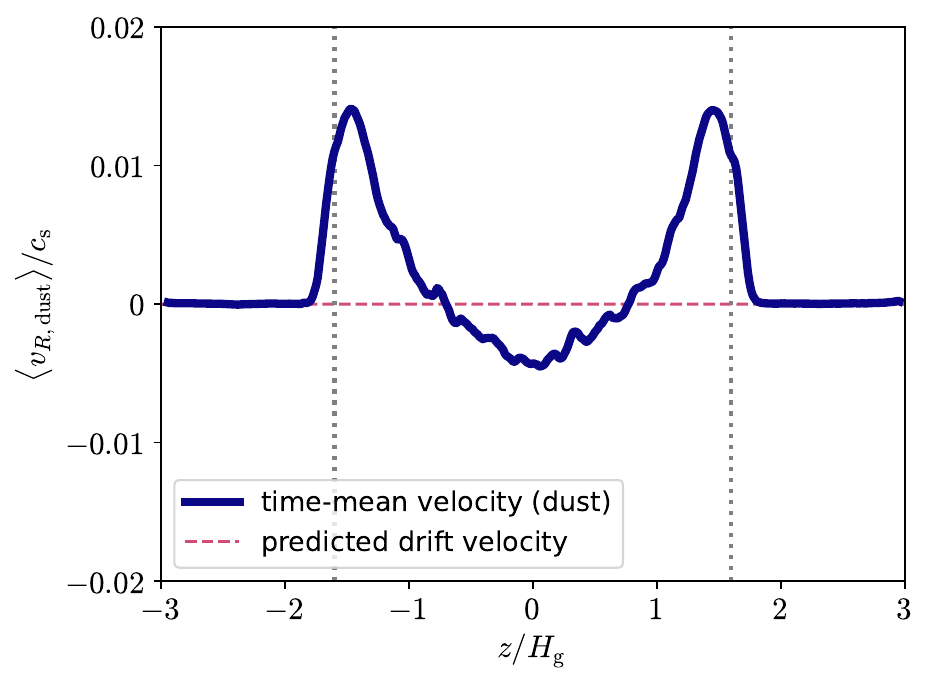}
    \end{center}
    \caption{Vertical profile of the time-averaged dust radial velocity $\langle v_{R,\rm dust}\rangle/c_{\rm s}$ at $R=40~{\rm au}$ for a run with $(Z_{\rm D2G},~a) = (0.01,~10~{\rm \mu m})$ in the dynamic-dust simulations. The dashed line indicates the predicted dust drift velocity $v_{\rm dust, drift}$ [Eq. \eqref{eq:dust_radial_drift_velocity}]. The dotted lines mark the height of $|z|=H_{\rm dg}$.}
    \label{fig:dust_radial_velocity}
\end{figure}  

In Sect. \ref{subsec:equilibrium_state_laminar}, we find the quasi-steady state for the dust vertical distribution and VSI-driven turbulence structure.
We also find that VSI-driven turbulence dominates dust radial motion in the dust layer.
Figure \ref{fig:dust_radial_velocity} plots the vertical profile of time mean dust radial velocity $\langle v_{R,\rm dust}\rangle$ for runs illustrated in Figs. \ref{fig:colormap_vz_rhodgINI_D2G1e-2_a1e-3_h} and \ref{fig:colormap_time_vzINI_rhodg_D2G1e-2_a1e-3_h}, where $v_{R,\rm dust} = v_{r,\rm dust}\sin{\theta}+v_{\theta,\rm dust}\cos{\theta}$ is the radial component of the dust velocity in the cylindrical coordinate.
The horizontal dashed line marks the radial drift velocity of dust due to gas drag, given by \citep{Whipple:1972vv,Adachi:1976uv,Weidenschilling:1977wt}
\begin{equation}\label{eq:dust_radial_drift_velocity}
    v_{\rm dust,drift} = -\frac{2{\rm St}}{1+{\rm St}^2}\eta v_{\rm K},
\end{equation}
where $\eta$ is the dimensionless quantity characterizing the pressure gradient of the gas and $v_{\rm K}=R\Omega_{\rm K}$ is the Keplerian velocity.
From the radial force balance, $\eta$ is given by
\begin{equation}
    \eta = -\frac{1}{2}\left(\frac{c_{\rm s}}{v_{\rm K}}\right)^2\frac{\mbox{d} \ln{P}}{\mbox{d} \ln{R}}.
\end{equation}
Our simulations assume the sub-Keplerian accretion disk with $\rho_{\rm gas} \propto R^p$ and $c_{\rm s} \propto R^{q/2}$, yielding $\eta = 0.01$ with $P = \rho_{\rm gas}c_{\rm s}^2/\gamma \propto R^{p+q}$ and $c_{\rm s}/v_{\rm K}=H_{\rm gas}/R = 0.1$ for $p=-1.0$ and $q=-1.0$ (see Sect. \ref{subsec:simulation_setup} and Table \ref{table:setup_parameter}).
Because the Stokes number takes values from $\sim 10^{-4}$ to $10^{-2}$ depending on the height, Eq. \eqref{eq:dust_radial_drift_velocity} indicate $v_{\rm dust, drift}\sim -10^{-6}$--$10^{-4} v_{\rm K}=-10^{-5}$--$10^{-3}c_{\rm s}$.
Figure \ref{fig:dust_radial_velocity} shows that in the dust layer ($|z| \lesssim 1.6H_{\rm gas}$), the dust moves not according to the radial drift velocity predicted by Eq. \eqref{eq:dust_radial_drift_velocity}.
Because the dust grain is small, this vertical profile of $\langle v_{R,\rm dust}\rangle$ traces that of the gas radial velocity.
This figure also indicates that the dust is in circulating motion because $\rho_{\rm dust}\langle v_{R,\rm dust}\rangle$ can take a similar value between at $z\approx0$ and $|z|\approx 1.6H_{\rm gas}$ (see Fig. \ref{fig:vertical_profile_rhod}).
This profile resembles previous simulations including dust particles (see Fig. 4 of \citealt{StollKley:2016vp}).

\section{Correlation timescale of VSI-driven turbulence}\label{appendix:correlation_timescale}
    \begin{figure}[t]
        \begin{center}
        \includegraphics[width=0.8\hsize,bb = 0 0 357 332]{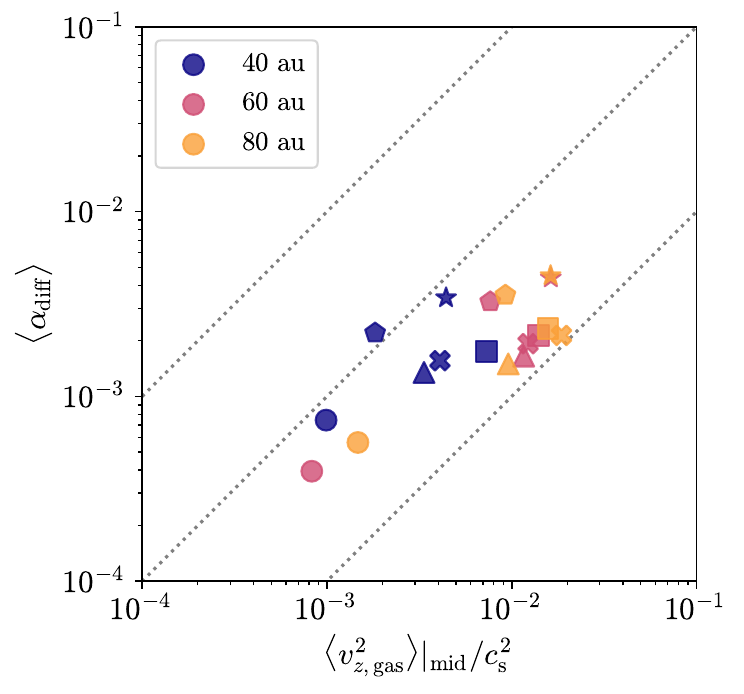}
        \end{center}
        \caption{Time mean squared vertical gas velocity at the midplane, $\langle v_{z,\rm gas}^2\rangle|_{\rm mid}$, vs. time mean dimensionless dust vertical diffusion coefficient, $\langle\alpha_{\rm diff}\rangle$, with different values of $R$ for runs of diffusion cases in dynamic-dust simulations. The symbols correspond to runs with $(Z_{\rm D2G},~a) = (0.003,~10~{\rm \mu m})$ (circles), $(0.01,~10~{\rm \mu m})$ (triangles), $(0.03,~10~{\rm \mu m})$ (crosses), $(0.1,~10~{\rm \mu m})$ (squares), $(0.03,~30~{\rm \mu m})$ (pentagons), and $(0.1,~30~{\rm \mu m})$ (stars). The dotted lines show Eq. \eqref{eq:relation_alphadiff_vz2} with $\tau_{\rm corr}\Omega_{\rm K} = 0.1,~1.0$, and $10.0$.}
        \label{fig:alpha_diff_vs_vz2_mid}
    \end{figure} 

    In Sect. \ref{subsec:equilibrium_state_laminar}, we estimate the dust vertical diffusion coefficient $\alpha_{\rm diff}$ from the dust layer's thickness.
    The independent measurements of $\alpha_{\rm diff}$ and squared vertical gas velocity $\langle v_{z, \rm gas}^2\rangle$ motivate us to estimate the correlation timescale of VSI-driven turbulence.
    To estimate the correlation time $\tau_{\rm corr}$, we introduce the relationship between $\alpha_{\rm diff}$ and $\langle v_{z, \rm gas}^2\rangle$, which is given by \citep{FromangPapaloizou:2006rz,YoudinLithwick2007}
    \begin{equation}\label{eq:relation_alphadiff_vz2}
        \alpha_{\rm diff} = \frac{\langle v_{z,\rm gas}^2 \rangle}{c_{\rm s}^2}\tau_{\rm corr}\Omega_{\rm K},
    \end{equation}
    where $\tau_{\rm corr}$ is the correlation timescale.
    
    In Fig. \ref{fig:alpha_diff_vs_vz2_mid}, we plot the time mean squared vertical gas velocity, $\langle v_{z,\rm gas}^2\rangle|_{\rm mid}$, versus time mean dust vertical diffusion coefficient, $\langle\alpha_{\rm diff}\rangle$, for the runs of diffusion cases in dynamic-dust simulations, with different values of $R$.
    This figure implies that the VSI-driven turbulence produces $0.1 \lesssim \tau_{\rm corr}\Omega_{\rm K} \lesssim 1.0$, which is an intermediate value between the values shown in previous studies ($\tau_{\rm corr}\Omega_{\rm K} \sim 0.2$ for \citealt{StollKley:2016vp}, and $\sim 20$ for \citealt{Flock:2020aa}).

\section{Cases of no VSI-driven turbulence in dynamic-dust simulations}\label{appendix:no_turbulence_case}

    \begin{figure}[t]
        \begin{center}
        \includegraphics[width=\hsize,bb = 0 0 434 340]{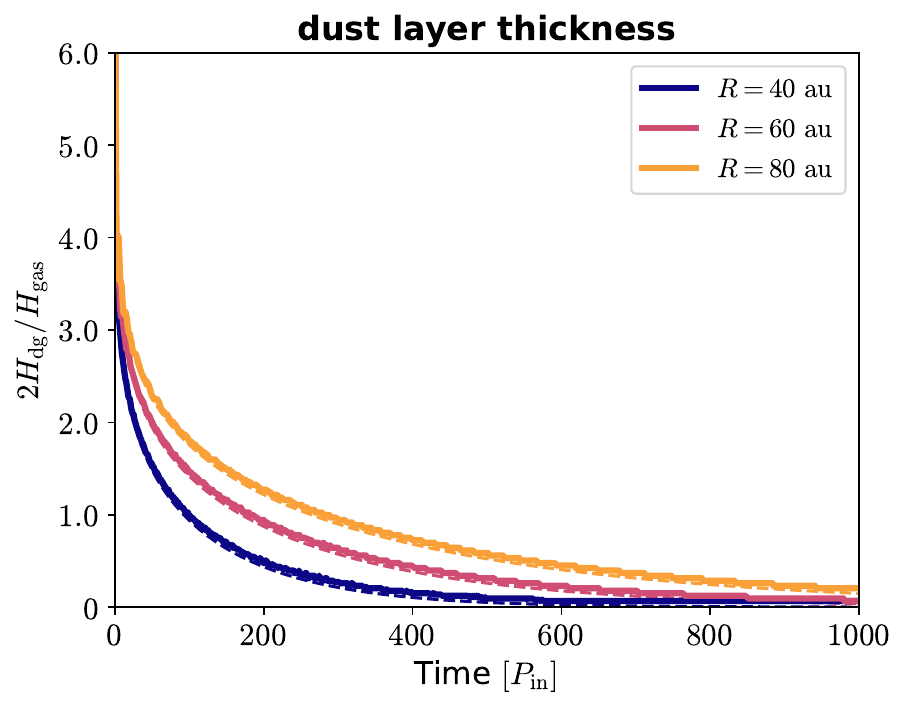}
        \end{center}
        \caption{Same as the upper panel of Fig. \ref{fig:2Hdg_alpha_TimeEvolution_D2G1e-2_a1e-3}, but for a run with $(Z_{\rm D2G},~a) = (0.01,~100~{\rm \mu m})$ in dynamic-dust simulations.}
        \label{fig:2Hdg_TimeEvolution_D2G1e-2_a1e-2}
    \end{figure}  

    In Sect. \ref{subsec:equilibrium_state_laminar}, for the dynamic-dust simulations, we find that the VSI-driven turbulence can not sustain the dust diffusion for the lower initial dust-to-gas mass ratios $Z_{\rm D2G}$ and larger dust grain's size $a$ (see Fig. \ref{fig:parameter_space_summary}).
    As an example of this case, Fig. \ref{fig:2Hdg_TimeEvolution_D2G1e-2_a1e-2} shows the time evolution of the dust layer thickness for a run with $(Z_{\rm D2G},~a) = (0.01,~100~{\rm \mu m})$.
    This figure indicates that the dust settles continuously toward the midplane due to the absence of VSI-driven turbulence.
    Because no turbulence operates, one expects that this run will never reach equilibrium and lead to a runaway dust setting.
    
\section{Cases of progress in fixed-to-dynamic-dust simulations}\label{appendix:progress_case}

    \begin{figure*}[t]
        \begin{center}
        \includegraphics[width=\hsize,bb = 0 0 751 528]{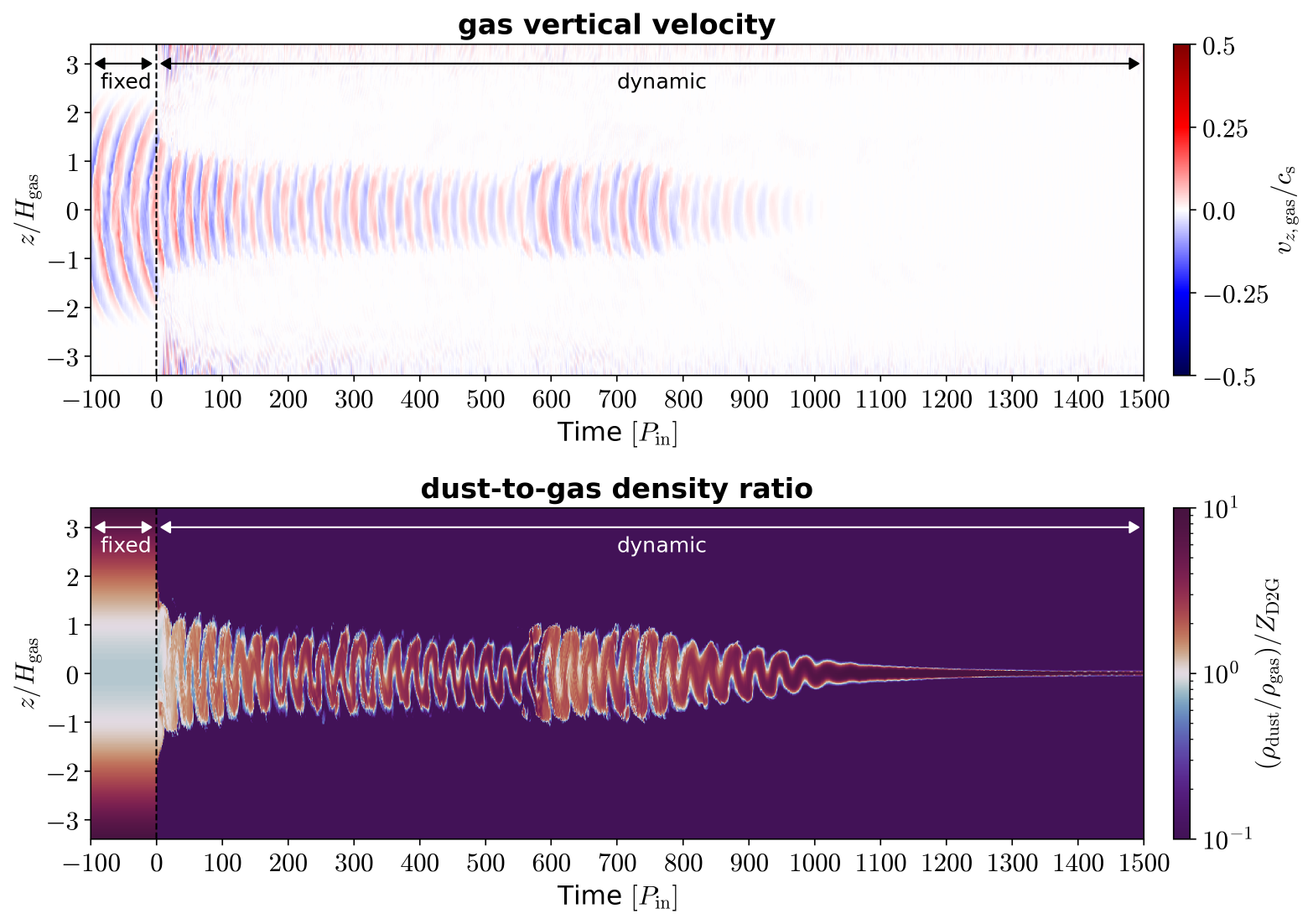}
        \end{center}
        \caption{Same as Fig. \ref{fig:colormap_time_vz_rhodgINI_turb_D2G1e-2_a1e-3_h}, but for a run with $(Z_{\rm D2G},~a) = (0.1,~100~{\rm \mu m})$ in fixed-to-dynamic-dust simulations.}
        \label{fig:colormap_time_vz_rhodgINI_turb_D2G1e-1_a1e-2_h_1500}
    \end{figure*}

    Figure \ref{fig:parameter_space_summary_turb} in Sect. \ref{subsec:equilibrium_state_turbulent} indicates that the run with $(Z_{\rm D2G},~a) = (0.1,~100~{\rm \mu m})$ in fixed-to-dynamic-dust simulations does not reach the equilibrium state within the simulated timeframe.
    We present in Fig. \ref{fig:colormap_time_vz_rhodgINI_turb_D2G1e-1_a1e-2_h_1500} the time evolution of gas vertical velocity $v_{z,\rm gas}$ and dust-to-gas density ratio $\rho_{\rm dust}/\rho_{\rm gas}$ at $R=40~{\rm au}$ for this run.
    During the fixed cooling timescale phase (i.e., until $t = 0$), VSI-driven turbulence develops in a manner similar to that seen in the run with $(Z_{\rm D2G},~a) = (0.01,~10~{\rm \mu m})$, as shown in Fig. \ref{fig:colormap_time_vz_rhodgINI_turb_D2G1e-2_a1e-3_h}.
    After switching to dynamic dust, VSI-driven turbulence gradually weakens over approximately $1000P_{\rm in}$.
    During this period, the dust layer becomes thinner and contracts toward the midplane, although some diffusion due to residual turbulence remains.
    Eventually, after about $1000P_{\rm in}$, the turbulence is fully suppressed, resulting in strong dust settling onto the midplane.

    \begin{figure*}[t]
        \begin{center}
        \includegraphics[width=\hsize,bb = 0 0 1264 288]{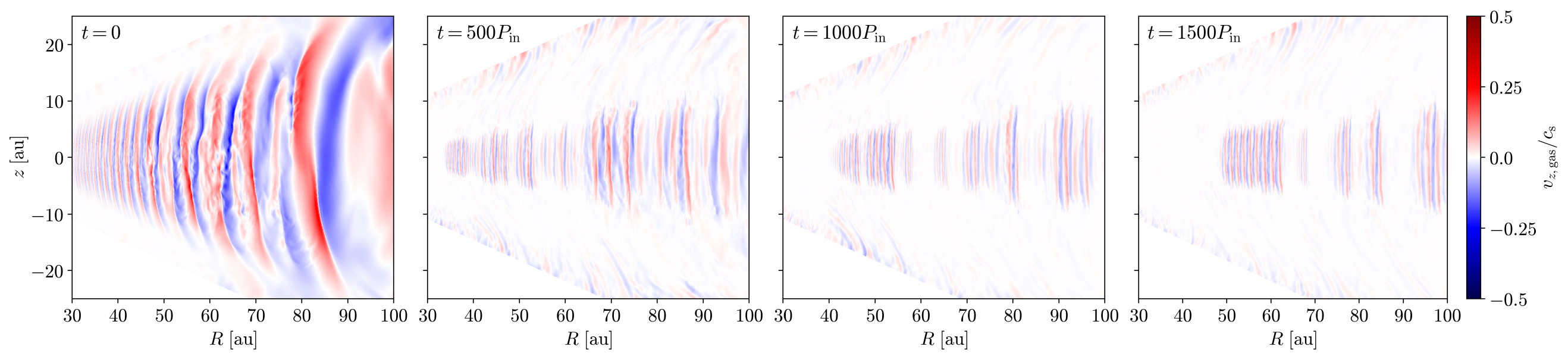}
        \end{center}
        \caption{Snapshots of gas vertical velocity $v_{z,\rm gas}/c_{\rm s}$ for the run displayed in Fig. \ref{fig:colormap_time_vz_rhodgINI_turb_D2G1e-1_a1e-2_h_1500}, as a function of $R$ and $z$ with different values of $t$.}
        \label{fig:colormap_vz_D2G1e-1_a1e-2_turb_time}
    \end{figure*}

    However, such a suppression of turbulence depends on the radial location.
    Figure \ref{fig:colormap_vz_D2G1e-1_a1e-2_turb_time} illustrates the spatial snapshots of the gas vertical velocity at different times for the run with $(Z_{\rm D2G},~a) = (0.01,~10~{\rm \mu m})$.
    The figure indicates that VSI-driven turbulence is confined to specific regions of the disk.
    For instance, a turbulence structure appears around $60~{\rm au}$ at $1500P_{\rm in}$, but not around $40~{\rm au}$.
    Moreover, as these turbulent regions gradually shrink over time, one can expect turbulence to vanish over the entire disk.

\end{appendix}

\end{document}